\documentclass{aa}  

\usepackage{graphicx}
\usepackage{txfonts}
\usepackage{lipsum}
\usepackage{subcaption}         
\usepackage{lscape}                                            
\usepackage{placeins}           
\usepackage{natbib}
\bibpunct{(}{)}{;}{a}{}{,} 

\begin{document}

   \title{Mock CSST NUV-NIR imaging of TNG50 galaxies with resolved stellar populations}

   \author{Jing Tang\inst{1}\corrauth{jtang@nao.cas.cn}       
        \and Ran Li\inst{2,1}
        \and Yang Chen\inst{3,1}
        \and Chao Liu\inst{1,4}
        \and Alessandro Bressan\inst{5,6}
        \and Léo Girardi\inst{7}
        \and Chengliang Wei\inst{8}
        \and Tianmeng Zhang\inst{1,4}
        }

   \institute{National Astronomical Observatories, Chinese Academy of Sciences, Beijing 100101, People’s Republic of China
   \and School of Physics and Astronomy, Beijing Normal University, Beijing 100875, China
   \and School of Physics, Anhui University, Hefei 230601, China
   \and School of Astronomy and Space Science, University of Chinese Academy of Sciences, Beijing 100049, People’s Republic of China
   \and SISSA, via Bonomea 365, I–34136 Trieste, Italy
   \and Nanjing Key Laboratory for Space and Underground Astrophysics, Nanjing 210023, China
   \and Osservatorio Astronomico di Padova – INAF, Vicolo dell’Osservatorio 5, 35122 Padova, Italy
   \and Purple Mountain Observatory, Chinese Academy of Sciences, Nanjing 210023, China
   }

   \date{Accepted October 28, 2026}
 
  \abstract 
   {}
   {We aim to develop a framework for generating synthetic images of resolved stellar populations from simulated galaxies and producing realistic CSST-like mock images, demonstrated here using two TNG50 galaxies as illustrative test cases.} 
   {Starting from simulated galaxies from TNG50, we use the population synthesis code TRILEGAL to sample individual stars from star particles, retaining bright stars explicitly and treating the remaining unresolved light, including faint stellar light and nebular emission from star-forming regions, as a diffuse background component of the galaxy. Based on the dust distribution inferred from the gas mass and metallicity of gas cells, we use the radiative transfer (RT) code SKIRT to construct the dust model. From this model, we compute the line-of-sight extinction for individual stars, while image-level transmission maps for the diffuse light are derived through RT simulations. For young stars still embedded in their natal molecular clouds, we additionally include birth-cloud attenuation. The resulting dust-attenuated stellar catalogues and diffuse-light images are then processed with the CSST simulation pipeline, which applies PSF convolution, sky background, and instrumental noise.}
   {Our framework produces mock CSST images for both a spiral galaxy and a dwarf irregular galaxy from TNG50 in the seven photometric bands. The mock images combine resolved stars, unresolved diffuse light, and dust attenuation within a unified framework, and include the main observational effects expected for the survey. The images are provided with known astrometric and photometric calibration information, and are intended as realistic image-level inputs for subsequent source detection, photometric recovery, and completeness analyses of resolved stellar populations. The framework is designed to be adaptable to other hydrodynamic simulations and observing configurations.}
   {}

   \keywords{methods: numerical --
                techniques: image processing --
                radiative transfer --
                ISM: dust, extinction --
                galaxies: stellar content
               }

   \maketitle
   \nolinenumbers

\section{Introduction}\label{sec:intro}
Resolved stellar population studies provide direct constraints on galaxy evolution by linking individual stars to spatially resolved star-formation and chemical-enrichment histories. To date, UV--optical resolved-star work beyond the Milky Way has been driven primarily by HST and its legacy programmes. HST's exquisite angular resolution has enabled landmark surveys across the Local Group and nearby galaxies, including the Panchromatic Hubble Andromeda Treasury in M31 (PHAT; \citealt{2012ApJS..200...18D}), its southern extension PHAST \citep{2025ApJ...979...35C}, PHATTER in M33 \citep{2021ApJS..253...53W}, the Hubble Tarantula Treasury Project in the LMC \citep{2013AJ....146...53S}, and nearby-galaxy programmes such as the Legacy ExtraGalactic UV Survey (LEGUS; \citealt{2015AJ....149...51C}), PHANGS–HST \citep{2022ApJS..258...10L}, and the Local Ultraviolet to Infrared Treasury (LUVIT; \citealt{2025ApJS..276....8G}). However, these observations typically cover limited areas or selected fields. 
In the coming decade, JWST, Euclid, and, later, Roman will greatly expand resolved-star capabilities in the near-infrared (NIR), with particular sensitivity to evolved and dust-obscured stellar populations. However, wide-field UV--optical mapping of young resolved stellar populations over large areas remains comparatively scarce. 

The upcoming Chinese Space-station Survey Telescope (CSST) is a 2 m-aperture space telescope operating in low-Earth orbit. Its main survey camera has a pixel scale of $\sim 0.074''\,\mathrm{pixel^{-1}}$ and a point-spread function (PSF) 80\% encircled-energy radius of $r_{80} = 0.15''$. The camera provides a field of view (FoV) of $1.1\,\mathrm{deg}^2$. Over its planned 10-year mission, CSST is designed to survey 17{,}500 $\mathrm{deg}^2$ of the sky, delivering multi-band imaging from the NUV to the NIR, with filters covering NUV (252--321 nm), u (321--401 nm), g (401--547 nm), r (547--692 nm), i (692--842 nm), z (842--1080 nm), and y (927--1080 nm). The expected 5$\sigma$ point-source limiting depths (AB) are $\sim$26 mag in g/r/i, $\sim$25 mag in NUV/u/z and $\sim$24.4 mag in y. In addition, its slitless spectroscopy offers three grisms (GU, GV, GI) with $R \simeq 200$ over 255--1000 nm, reaching broadband-equivalent depths of $\sim$23 mag \citep{Zhan2021}.

With this expected optical depth, CSST should be able to detect the brightest blue-supergiant candidates in nearby galaxies out to $\sim 40\,\mathrm{Mpc}$, adopting a conservative $M_{\mathrm{opt}} \simeq -7$ mag and neglecting extinction. This is well within the luminosity range of late-B and early-A supergiants, which can reach $M_V \simeq -9.5$ mag \citep{2008ApJ...681..269K}. A simple volume extrapolation from the 1,678 galaxies currently known within 11 Mpc\footnote{\url{http://www.sao.ru/lv/lvgdb/introduction.php}} implies a potential parent sample of order $10^{5}$
galaxies, although the effective resolved-star sample will depend on distance, crowding, background level, dust attenuation, and the survey footprint. Compared with HST programmes that often target limited fields, the wide FoV of CSST will enable contiguous, galaxy-wide maps of resolved luminous stars for many nearby galaxies, with uniform multi-band photometry and slitless spectroscopy where feasible. These data will provide spatially comprehensive constraints on star-formation histories (SFHs), stellar-population gradients, and dust attenuation across galaxy disks and, where depth and crowding permit, into central regions and outer halos. The NUV coverage is particularly valuable for identifying young massive stars, tracing recent star formation, and constraining dust attenuation in young stellar populations. Together, these capabilities will extend resolved stellar population studies from localised case studies to more uniform, statistically controlled, galaxy-scale analyses of nearby galaxies.

Resolved stellar population studies with CSST will critically depend on reliable source detection, photometry, and stellar-parameter inference, all of which are challenging in realistic images of nearby galaxies. In crowded fields, source confusion and blending reduce completeness and degrade photometric accuracy. Spatial variations in the contrast between stars and the local background, set by unresolved stellar light, nebular emission, and sky background, produce position-dependent detection limits and systematic flux biases. Dust attenuation further complicates the problem: patchy extinction from the diffuse interstellar medium (ISM) alters colours and suppresses fluxes along different sightlines, while young stars may experience additional localised attenuation from their birth clouds, introducing age-dependent selection effects. These effects bias the recovered luminosities and colours and, consequently, the inferred attenuation and stellar properties. This motivates realistic, truth-matched mock observations that include crowding, complex image backgrounds, physically motivated dust attenuation, and the dominant image-level observational effects. In this work, we construct synthetic galaxy images and forward-model them through the CSST simulation pipeline, applying PSF convolution, sky background, and instrumental noise, to produce CSST-like mock images.

Hydrodynamical cosmological simulations provide a natural basis for such mocks. Existing efforts broadly fall into two categories. Image-level RT calculations, for example those performed with SKIRT, enable comparisons with imaging surveys in integrated-light morphology and colours \citep[e.g.][]{2019MNRAS.483.4140R,2024A&A...683A.181B}, but typically do not provide truth-matched catalogues of individually resolved stars. Conversely, catalogue-level products, often constructed from Milky Way (MW) analogues in simulations such as Auriga and FIRE, are well suited to resolved-star diagnostics, but do not by themselves produce realistic images and therefore do not capture PSF blending, complex image backgrounds, or image-level noise. In many catalogue-based applications, dust attenuation is treated with simplified prescriptions, including external MW extinction maps \citep{2018MNRAS.481.1726G}, distance-dependent laws \citep{2018MNRAS.480..722V}, or per-star extinctions inferred from line-of-sight gas columns mapped to $E(B-V)$ \citep{2020ApJS..246....6S}. These prescriptions are useful for MW studies, but do not straightforwardly generalise to external galaxies with different dust compositions, geometries, and viewing angles.

Here we present an image-level forward-modelling framework based on the TNG50 simulation that bridges these approaches. The framework combines (i) a truth-matched catalogue of resolved stars, (ii) an unresolved diffuse-light image, and (iii) geometry- and wavelength-dependent SKIRT-based dust attenuation. The resolved stars and unresolved diffuse light are attenuated separately. We demonstrate the framework using one spiral galaxy and one dwarf irregular galaxy from TNG50. These two systems serve as illustrative test cases rather than a systematic exploration of galaxy properties and observing configurations.

This paper represents a first step towards preparing resolved stellar population studies of nearby galaxies with CSST, focusing on the construction of physically motivated CSST-like mock images. The organisation of the paper is as follows: Sect.~\ref{sec:methods} describes the simulation data and the tools employed. Sect.~\ref{sec:synthetic} details the construction of the resolved stellar catalogues, the assignment of per-star extinction, and the treatment of attenuated unresolved diffuse light. Sect.~\ref{sec:dust_effect} estimates the impact of dust attenuation on resolved stellar populations. Sect.~\ref{sec:mock} presents the production of mock CSST images. Finally, Sect.~\ref{sec:summary} summarises our results and outlines future extensions.

\section{Simulation data and tools}\label{sec:methods}
\subsection{The TNG50 simulation}
TNG50 \citep{2019MNRAS.490.3196P,2019MNRAS.490.3234N} is the highest-resolution run of the IllustrisTNG project \citep{2018MNRAS.480.5113M,2018MNRAS.477.1206N,2018MNRAS.475..624N,2018MNRAS.475..648P,2018MNRAS.475..676S}, a suite of cosmological magneto-hydrodynamic simulations performed with the moving-mesh code AREPO \citep{2010MNRAS.401..791S}. It follows the coupled evolution of dark matter (DM), gas, stars, black holes and magnetic fields within a cubic volume of 51.7 comoving Mpc on a side, initialised with $(2160)^{3}$ DM particles and $(2160)^{3}$ gas cells. Its baryonic mass resolution is $8.5 \times 10^{4}\,\mathrm{M_\odot}$, and the average cell size in the star-forming regions of galaxies is 70--140 pc, comparable to that of typical zoom-in simulations. The cosmological parameters adopted in the simulation are $\Omega_{m} = 0.3089$, $\Omega_{b} = 0.0486$, $\Omega_{\Lambda} = 0.6911$, $H_0 = 100\,h\,\mathrm{km\,s^{-1}\,Mpc^{-1}}$ with $h = 0.6774$, $\sigma_{8} = 0.8159$, and $n_{s} = 0.9667$, consistent with the Planck 2015 results \citep{2016A&A...594A..13P}. For the galaxy formation model \citep{2017MNRAS.465.3291W,2018MNRAS.473.4077P}, a  \citet{2003PASP..115..763C} initial mass function (IMF) is assumed for star particles in galaxies.

\subsection{The population synthesis code TRILEGAL}\label{sec:trilegal}
The TRILEGAL code \citep{2005A&A...436..895G} is a tool for generating synthetic stellar catalogues for both the MW and external galaxies \citep{2007ASPC..378...20G}. In particular, it has been used to construct the first comprehensive MW mock stellar catalogue for the CSST main survey camera \citep{2023SCPMA..6619511C}. It is built on four key ingredients:
\begin{itemize}
\item a library of stellar evolutionary tracks, precomputed into isochrones;
\item a library of synthetic spectra, preprocessed into bolometric-correction (BC) tables;
\item a set of filter, detector, and telescope throughput curves, which define the photometric system;
\item a galaxy description specified by the IMF, the SFH, and the age-metallicity relation (AMR).
\end{itemize}

Given these inputs, TRILEGAL performs a Monte Carlo (MC) simulation to sample
individual stars from the prescribed stellar population model. It then interpolates within the isochrone grids and applies the BC tables to compute magnitudes in the chosen set of passbands.

\begin{figure}
\centering
\includegraphics[width=0.85\hsize]{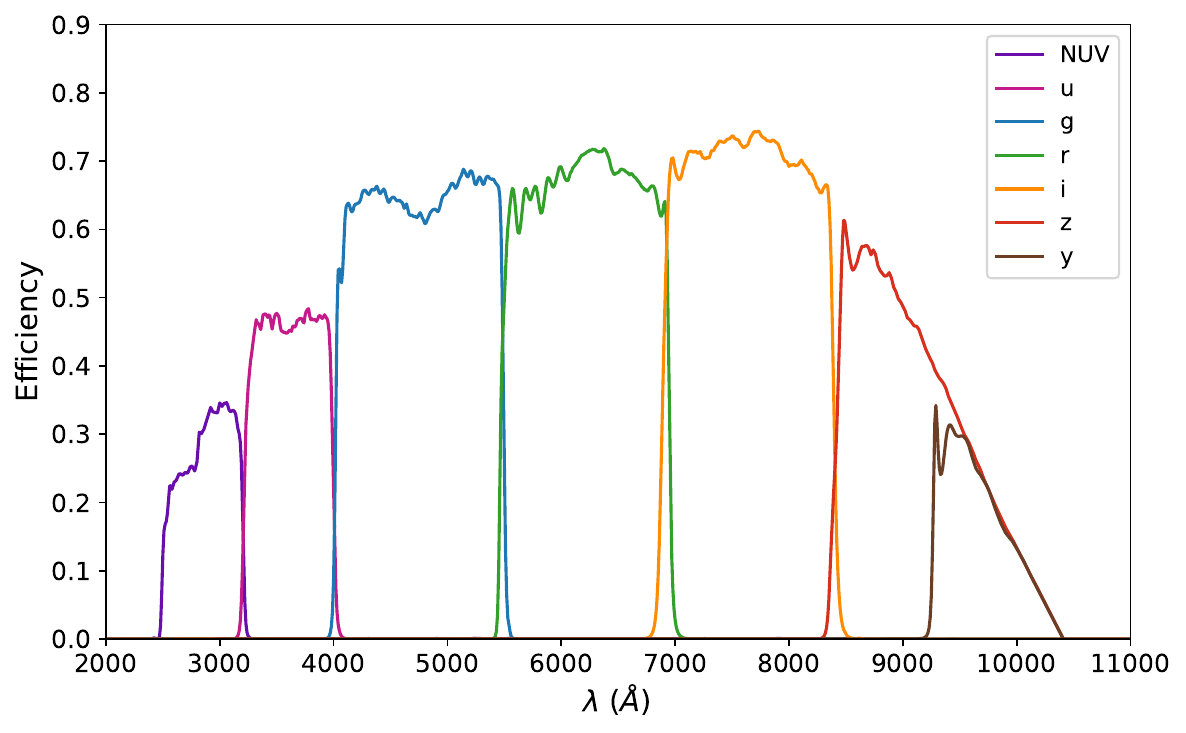}
\caption{The system throughput curves for CSST main survey camera imaging bands. These curves are multiplicative products of three components: telescope mirror efficiencies, filter transmissions and detector quantum efficiencies. The curves are taken from the official CSST simulation repository.}
\label{fig1}
\end{figure}

In this work, the input datasets adopted for TRILEGAL are as follows. For stellar evolution models, we use PARSEC v1.2S \citep{2012MNRAS.427..127B,2014MNRAS.444.2525C,2014MNRAS.445.4287T,2015MNRAS.452.1068C}, complemented by COLIBRI tracks for TP-AGB phase \citep{2017ApJ...835...77M,2019MNRAS.485.5666P,2020MNRAS.498.3283P}. 
The BC tables are computed with the built-in YBC code \citep{2019A&A...632A.105C}, and magnitudes are calculated using the CSST bandpass throughput curves\footnote{\url{https://csst-tb.bao.ac.cn/code/csst-sims/csst_msc_sim/-/tree/v3.3.0/observation_sim/instruments/data/throughputs}} (as shown in Fig.~\ref{fig1}). Although YBC allows for extinction-dependent BCs, we use
unextincted BC tables; dust attenuation is applied subsequently on
a star-by-star basis, as described in Sect.~\ref{sec:attenuation}. 
For the galaxy model, instead of adopting analytic parameterisations for the SFH and AMR, we take these quantities directly from TNG50: each simulated star particle provides an initial stellar mass, an age and a metallicity. For the IMF, we adopt the same choice as in TNG50, namely the \citet{2003PASP..115..763C} IMF, in the form of
\begin{equation}
\label{eq1}
\xi(m) \propto
\begin{cases}
m^{-1}\exp\!\left[-\dfrac{\bigl(\log\,m-\log\,m_c\bigr)^2}{2\sigma^2}\right], & m \le 1\,\mathrm{M_\odot}, \\[6pt]
m^{-2.3}, & m > 1\,\mathrm{M_\odot},
\end{cases}
\end{equation}
where $m_c = 0.079$ and $\sigma = 0.69$. The normalisation constants are determined in TRILEGAL by enforcing continuity at the transition mass and by requiring the IMF to be mass-normalised, i.e. $\int m\,\xi_m\,dm = 1\,\mathrm{M_\odot}$ over the adopted mass range from $0.1\,\mathrm{M_\odot}$ to $100\,\mathrm{M_\odot}$.

For binaries, we follow the default settings in TRILEGAL, assuming a binary fraction of 30\% and a flat mass-ratio distribution over $q \in [0.7, 1]$ \citep{2002A&A...385..847B}. Here, ``binaries'' refers to non-interacting systems. To avoid generating an excessive number of faint stars, we retain only stars with $M \geq 0.7\,\mathrm{M_\odot}$. In the TRILEGAL run, we fix the distance to 10 pc to obtain absolute magnitudes and neglect foreground extinction.

\subsection{The radiative transfer code SKIRT}\label{sec:skirt}
SKIRT is a 3D Monte Carlo (MC) radiative transfer (RT) code \citep{2011ApJS..196...22B,2015A&C.....9...20C,2020A&C....3100381C}. It represents radiation as MC photon packets and tracks their propagation through dust, accounting for anisotropic scattering, absorption and (re-)emission, and producing synthetic observations over user-defined wavelength grids. SKIRT is equipped with an extensive library of SEDs, dust models, geometry prescriptions and spatial grid schemes, together with throughput curves for a variety of instruments.

In this work, we use the released version 9.0 of SKIRT, and adopt a configuration that largely follows previous TNG50 post-processing studies \citep{2022MNRAS.516.3728T,2024A&A...683A.181B}. Below we describe the prescriptions for the two key components of our SKIRT simulations: the primary radiation sources and the diffuse dust medium.

\subsubsection{Primary radiation sources}\label{sec:radiation}
For each simulated galaxy, the radiation sources are represented by star particles. We compute their emission using an age-dependent SED treatment.

Evolved star particles: Particles older than 10\,Myr---the typical lifetime of a H\,{\sc ii} region---are assigned stellar population SEDs from the \citet[][hereafter BC03]{2003MNRAS.344.1000B} library with a \citet{2003PASP..115..763C} IMF. The template inputs, including initial mass $M_\mathrm{init}$, metallicity $Z$ and age $t$, are taken directly from the simulation. 

Young star particles: Particles younger than 10\,Myr are assumed to remain embedded in their natal molecular clouds and are therefore modelled with MAPPINGS III SEDs \citep{2008ApJS..176..438G}. The MAPPINGS III templates are specified by five parameters: 
\begin{itemize}
\item Star formation rate (SFR): calculated as the particle's initial mass $M_\mathrm{init}$ divided by 10 Myr, consistent with the templates' assumption of constant star formation over the past 10 Myr.

\item Metallicity $Z$: taken directly from the simulation. 

\item H\,{\sc ii} region compactness parameter $\mathcal{C}$: a dimensionless parameter that quantifies the compactness of the star-forming region and is closely related to the radiation-field intensity at the ionisation front of the H\,{\sc ii} region \citep{2008ApJS..176..438G}. Larger values of $C$ imply stronger dust heating, leading to warmer dust and shifting the far-infrared (FIR) dust re-emission spectrum to shorter wavelengths. Following \citet{2021MNRAS.506.5703K}, we sample log\,$\mathcal{C}$ from a Gaussian distribution with mean 5 and standard deviation 0.4. 

\item ISM pressure $P_0$: the pressure of the diffuse ISM surrounding the
H\,{\sc ii} region, which affects the FIR dust re-emission spectrum. Following \citet{2024A&A...683A.181B}, we adopt a fixed value of $P_0 = 1.38 \times 10^{-12}\,\mathrm{Pa}$, where $\mathrm{Pa}$ denotes Pascal.  

\item PDR covering factor $f_\mathrm{PDR}$: quantifies the degree to which H\,{\sc ii} regions are enveloped by photodissociation regions (PDRs). It has a significant effect on the whole SED, and it is set to 
\begin{equation}
\label{eq2}
f_\mathrm{PDR} = e^{-t/\tau_\mathrm{clear}},
\end{equation}
where $t$ is the particle age and $\tau_\mathrm{clear}$ is the PDR clearing timescale. It means the PDR coverage decreases as the H\,{\sc ii} region evolves. Here we adopt $\tau_\mathrm{clear} = 3$ Myr \citep{2022MNRAS.516.3728T}. 
\end{itemize}

The geometry of the radiation field is determined by star-particle positions from the simulation and by their smoothing lengths, $h_\mathrm{sml}$, defined as follows. For evolved particles, $h_\mathrm{sml}$ is taken directly from the simulation as the 3D distance to the $(32\pm1)^\mathrm{th}$ nearest star particle. For young particles, instead of using the simulation-based smoothing length, we assign an effective radius $R$ to each particle based on its SFR. This radius is obtained by inverting the SFR--radius relation fitted to the SINGS
galaxy sample of \citet{2003PASP..115..928K} \citep[see][Figure~8]{2015MNRAS.450.1812L},
\begin{equation}
\label{eq3}
\mathrm{SFR}\,(\mathrm{M_\odot\,yr}^{-1}) 
= 9.58 \times 10^{-8}\,\left(\frac{R}{\mathrm{pc}}\right)^2, 
\end{equation}
We adopt this effective radius as the smoothing length, i.e. $h_{\rm sml}=R$. Thus, each young particle is interpreted as an unresolved star-forming complex at the resolution of TNG50. The luminosity of each star particle, computed from the stellar population SED, is then distributed using the smoothed particle hydrodynamics (SPH) spline kernel defined on [0, $h_\mathrm{sml}$] \citep{2001NewA....6...79S}: 
\begin{equation}
\label{eq4}
W(r,h_\mathrm{sml})=\frac{2^{
v}\sigma}{h_\mathrm{sml}^{
v}}
\begin{cases}
1-6q^{2}+6q^{3}, & 0\le q \le \frac{1}{2},\\[4pt]
2(1-q)^{3},      & \frac{1}{2} < q \le 1,\\[4pt]
0,               & q>1,
\end{cases}
\end{equation}
where $q \equiv r/h_\mathrm{sml}$, $r$ is the radial distance from the particle centre, $v$ denotes the number of spatial dimensions in which the
kernel is normalised. The normalisation constant is $\sigma=1/\pi$ for the 3D kernel ($v=3$). In the SKIRT calculations, the 3D form is adopted to distribute stellar-particle luminosities.

\subsubsection{Diffuse dust medium}\label{sec:dust}
Since dust physics is not explicitly modelled in TNG50, we infer the  interstellar dust content of the simulated galaxy from gas cells. We first distinguish the cool interstellar gas from the hotter circumgalactic
gas using the temperature-density criterion \citep{2012MNRAS.427.2224T,2019MNRAS.484.5587T},
\begin{equation}
\label{eq5}
\log\,\left(\frac{T_\mathrm{gas}}{\mathrm K}\right) < 6+0.25\,\log\,\left(\frac{\rho_\mathrm{gas}}{10^{10}\,\mathrm{M_\odot}\,h^2\,\mathrm {kpc}^{-3}}\right),
\end{equation}
where $T_\mathrm{gas}$ and $\rho_\mathrm{gas}$ are the gas temperature and mass density, respectively. Gas cells satisfying Eq.~(\ref{eq5}) are assumed to be eligible to host dust, and their dust mass is then computed as 
\begin{equation}
\label{eq6}
\mathrm{M_{dust}} = f_\mathrm{dust}\,Z_\mathrm{gas}\,\mathrm{M_{gas}},
\end{equation}
where $Z_\mathrm{gas}$ is the gas metallicity, $\mathrm{M_{gas}}$ is the gas mass, and $f_\mathrm{dust}$ is the dust-to-metal ratio, representing the fraction of metals locked up in dust grains (rather than remaining in the gas phase). Here we adopt $f_\mathrm{dust} = 0.2$, together with the previously introduced  $\tau_\mathrm{clear} = 3$ Myr, jointly calibrated by comparing TNG50 with observations of nearby galaxies \citep{2022MNRAS.516.3728T}. 

The dust geometry is specified by the positions and smoothing lengths of the gas cells that contain dust, both taken directly from the simulation. For each cell, the smoothing length is defined as the radius of a sphere centred on the cell that encloses the 64 $\pm$ 1 nearest DM particles, which differs from the definition adopted for star particles. The dust density field is then constructed by convolving the cell dust masses with the 3D SPH kernel (Eq.~(\ref{eq4})). For the dust model, we adopt THEMIS \citep{2017A&A...602A..46J}, in which dust grains are represented as a mixture of amorphous silicates and hydrogenated amorphous carbons. We note that different models, such as THEMIS and the \citet{2004ApJS..152..211Z} model, adopt different grain compositions and size distributions, and therefore predict different mass extinction coefficients. In addition, alternative prescriptions for inferring dust masses, beyond that adopted in Eq.~(\ref{eq5}), can lead to different dust density fields. Examples include prescriptions based on different temperature cutoffs or dust-to-metal ratios \citep{2016MNRAS.462.1057C,2018ApJS..234...20C,2017MNRAS.470..771T,2019MNRAS.487.1844M,2020MNRAS.497.4773S,2020MNRAS.492.5167V,2022MNRAS.510.3321P}. Both the choice of dust model and the dust-mass prescription can therefore affect the predicted extinction.

\subsubsection{Additional SKIRT configuration}\label{sec:skirt_config}
We consider primary radiation sources over the wavelength range 0.09--1.2 $\mu \mathrm{m}$, and adopt $2 \times 10^{9}$ photon packets in each simulation as a compromise between signal-to-noise (S/N) ratio and computational cost. An adaptive octree grid is used to reconstruct the dust density field, with refinement parameters varying among individual galaxies. The grain size distribution is discretised into 15 bins for both the silicate and hydrocarbon dust populations. The final output images have a fixed size of $800 \times 800$ pixels.

For each galaxy, we produce dust-attenuated and transparent total-light
images in the CSST bands with the RT simulations. Both images are generated through photon-packet propagation and are used in Sect.~\ref{sec:dl} to construct image-level transmission maps. In addition, we use SKIRT probes to output transparent, noise-free maps of the imported stellar radiation sources. These probe maps are direct
projections of the input source luminosities, computed from the assigned
stellar population SEDs, smoothed with the 3D SPH kernel, and projected onto
the image plane. They do not involve photon-packet propagation and are
therefore not affected by MC noise. We output these probe maps
separately for the recent ($\leq 100\,\mathrm{Myr}$) and older
($>100\,\mathrm{Myr}$) stellar components. They are later used in Sect.~\ref{sec:dl} to construct the unresolved diffuse light; this 100\,Myr split is distinct from the 10\,Myr young/evolved division adopted for the SED
assignment. All images are projected onto the $(x_\mathrm{p}, y_\mathrm{p})$ plane of the viewing coordinate system, i.e. the image plane, using the transformation
\begin{equation}
\label{eq7}
\begin{pmatrix}
x_\mathrm{p}\\
y_\mathrm{p}\\
z_\mathrm{p}
\end{pmatrix}
=
\begin{pmatrix}
0 & 1 & 0\\
-1 & 0 & 0\\
0 & 0 & 1
\end{pmatrix}
\begin{pmatrix}
x\\
y\\
z
\end{pmatrix},
\end{equation}
where $(x, y, z)$ denotes the original TNG50 coordinates.

\section{Synthetic galaxy products}\label{sec:synthetic}
\subsection{Showcase galaxies from TNG50}
To limit the data volume and computation time, in this work we select two star-forming galaxies from TNG50 with stellar masses below $10^{10}~\mathrm{M_\odot}$---one spiral galaxy and one dwarf irregular galaxy---as case studies to illustrate the workflow. Their basic properties are summarised in Table~\ref{table:1}.

For each galaxy, we use the $z=0$ snapshot of the TNG50 simulation and extract
the associated star particles within $5\,r_{1/2,\star}$ and gas cells within
$10\,r_{1/2,\star}$, where $r_{1/2,\star}$ is the stellar half-mass radius. The same procedure is applied to both
galaxies, while several galaxy-dependent setup parameters, including the
adopted distance, image FoV, pixel scale, octree domain, and dust-grid
refinement settings, are adjusted to account for differences in physical size
and gas morphology, as listed in Table~\ref{table:2}. To validate the dust-grid
construction, we compare the input dust distributions with their octree
reconstructions in Appendix~A.
\begin{table*}
\caption {Main properties of the two selected TNG50 galaxies.}
\label{table:1} 
\centering
\tabcolsep 15pt
\begin{tabular}{cccccccc}
\hline\hline 
ID & $M_{\star}$ & $M_\mathrm{gas}$ & $Z_{\star} (2r_{1/2,\star})$ & $Z_\mathrm{gas} (2r_{1/2,\star})$ & $r_{1/2,\star}$ & SFR & $N_{\mathrm{young}}/N_{\mathrm{total}}$\\ 
& ($\mathrm{M_\odot}$) & ($\mathrm{M_\odot}$) & & & (kpc) & ($\mathrm{M_\odot\,yr}^{-1}$) \\
\hline
666210 & 7.2E09 & 2.1E10 & 0.013 & 0.013 & 8.3 & 0.89 & 93/118173\\  
860672 & 5.3E07 & 6.4E08 & 0.006 & 0.007 & 1.1 & 0.0045 & 5/913\\  
\hline
\end{tabular}
\tablefoot{The columns are (left to right) (1) Subhalo ID; (2) Total mass of all star particles in this subhalo; (3) Total mass of all gas cells in this subhalo; (4) Mass-weighted average metallicity of the star particles within twice the stellar half-mass radius; (5) Mass-weighted average metallicity of the gas cells within twice the stellar half-mass radius; (6) Stellar half-mass radius; (7) Sum of the individual SFRs of all gas cells in this subhalo; (8) Number of young star particles ($\le$ 10 Myr) out of total star particles within $5\,r_{1/2,\star}$.}
\end{table*}

\begin{table}
\centering
\caption{Galaxy-dependent configuration in SKIRT for synthetic image generation.}
\label{table:2}
\begin{tabular}{lcc}
\hline
Parameter & TNG666210 & TNG860672 \\
\hline
Distance [Mpc] & 10 & 1 \\
FoV [kpc] & 40 & 4 \\
Pixel scale [pc pixel$^{-1}$] & 50 & 5 \\
Octree box [kpc] & $[-83,83]^3$ & $[-11,11]^3$ \\
\texttt{minLevel} & 6 & 6 \\
\texttt{maxLevel} & 14 & 12 \\
\texttt{maxDustFraction} & 1e-6 & 1e-6 \\
\texttt{maxDustOpticalDepth} & 0.01 & 0 \\
\hline
\end{tabular}
\tablefoot{The octree box defines the 3D spatial domain of the dust grid, with $x,y,z\in[a,b]$. 
\texttt{minLevel} and \texttt{maxLevel} specify the minimum and maximum octree refinement levels, respectively. \texttt{maxDustFraction} is the maximum allowed fraction of the total dust mass in any individual grid cell. \texttt{maxDustOpticalDepth} is the maximum allowed dust optical depth along a cell diagonal at $0.55\,\mu\mathrm{m}$. For TNG860672, a value of zero disables this refinement criterion.}
\end{table}

\subsection{Resolved stellar catalogues}\label{sec:catalogue}
We use TRILEGAL (Sect.~\ref{sec:trilegal}) to resolve star particles into individual stars. 

Evolved star particles: Each particle is treated as a simple stellar population (SSP). The particle initial mass, age, and metallicity from the simulation are passed to TRILEGAL, which generates stars spanning a range of masses at fixed age and metallicity. TRILEGAL provides the absolute magnitudes of these stars, which are subsequently converted to apparent magnitudes based on the adopted galaxy distance. 

The spatial positions of the resolved stars are assigned statistically. For each parent particle, we construct a local multivariate normal distribution centred on the particle position, with the covariance matrix estimated from the positions of neighbouring star particles within its $h_\mathrm{sml}$, as defined in Sect.~\ref{sec:radiation}. The resolved stars associated with that parent particle are then randomly drawn from this distribution.

Young star particles: Each particle is treated as a composite stellar population, assuming continuous star formation at a constant rate over the past 10 Myr. The particle initial mass and metallicity are taken from the simulation and passed to TRILEGAL. TRILEGAL then samples stars spanning a range of masses and ages, with ages drawn on a logarithmic grid from $\mathrm{log}\,(t/\mathrm{yr})$ = 5.05 to 6.99 in steps of 0.02.

The spatial distribution of these stars is assigned in a simplified statistical manner. Although star formation is expected to be hierarchical, with stars forming in clustered substructures within larger star-forming regions, such complexity is beyond the scope of this work. Instead, we distribute the stars associated with each young particle using a 3D isotropic Gaussian centred on the particle position, with
\begin{equation}
\label{eq8}
\sigma = \frac{h_\mathrm{sml}}{3}f,
\end{equation}
where $h_\mathrm{sml}$ is the smoothing length defined in Eq.~(\ref{eq3}), and $f$ is a random factor uniformly drawn from $[0.9, 1.1]$. This treatment is not intended to reproduce the detailed internal structure of star-forming regions, but is sufficient for constructing mock images and testing the stellar detection pipeline.

We adopt a magnitude cut to define the practical boundary between stars treated explicitly as point sources and the remaining unresolved stellar light. For both example galaxies, stars are retained if they satisfy $m < m_\mathrm{lim} + 1$ mag in at least one of the seven CSST bands, where $m_\mathrm{lim}$ denotes the nominal 5$\sigma$ point-source limiting magnitude for the two-exposure CSST depth. This fiducial cut corresponds to about 2$\sigma$ in the coadded depth, or about 1.4$\sigma$ in a single 150 s exposure. We emphasise that this is not intended as a source-detection threshold, but as a practical choice for retaining faint stellar granularity in the mock images. Stars fainter than this cut are included in the unresolved diffuse-light component described in Sect.~\ref{sec:dl}. 

To assess the sensitivity to this choice, we repeat the decomposition of TNG860672 using the shallower \(m<m_{\rm lim}\) criterion and compare it with the fiducial $m < m_\mathrm{lim} + 1$ case. Adopting the deeper magnitude cut increases the number of stars retained from $4.13\times10^5$ to $7.67\times10^5$. Using the i band as a representative case, the diffuse-light flux decreases by about 20\%, with the corresponding amount of stellar flux reassigned to explicitly rendered point sources. We quantify the associated change in the small-scale diffuse background using the diffuse-light images before the clipping step described in Sect.~\ref{sec:dl}. For each image, we construct a high-pass residual by subtracting a Gaussian-smoothed version from the original image. We repeat this calculation using Gaussian kernels with full widths at half maximum of 4–16 native pixels (20–80 pc). With the deeper cut, the root-mean-square amplitude of the high-pass residual decreases by about 4\% across these scales, while the high-pass residuals obtained with the two magnitude cuts remain highly correlated, with a pixel-by-pixel Pearson correlation coefficient of $r\simeq0.997$. For this representative case, the test therefore indicates that, despite the change in the partition of stellar light between the point-source and diffuse components, the small-scale fluctuation amplitude and spatial pattern of the diffuse background are weakly sensitive to the adopted magnitude cut.

\begin{figure}
    \centering
    {\includegraphics[width=0.85\hsize]{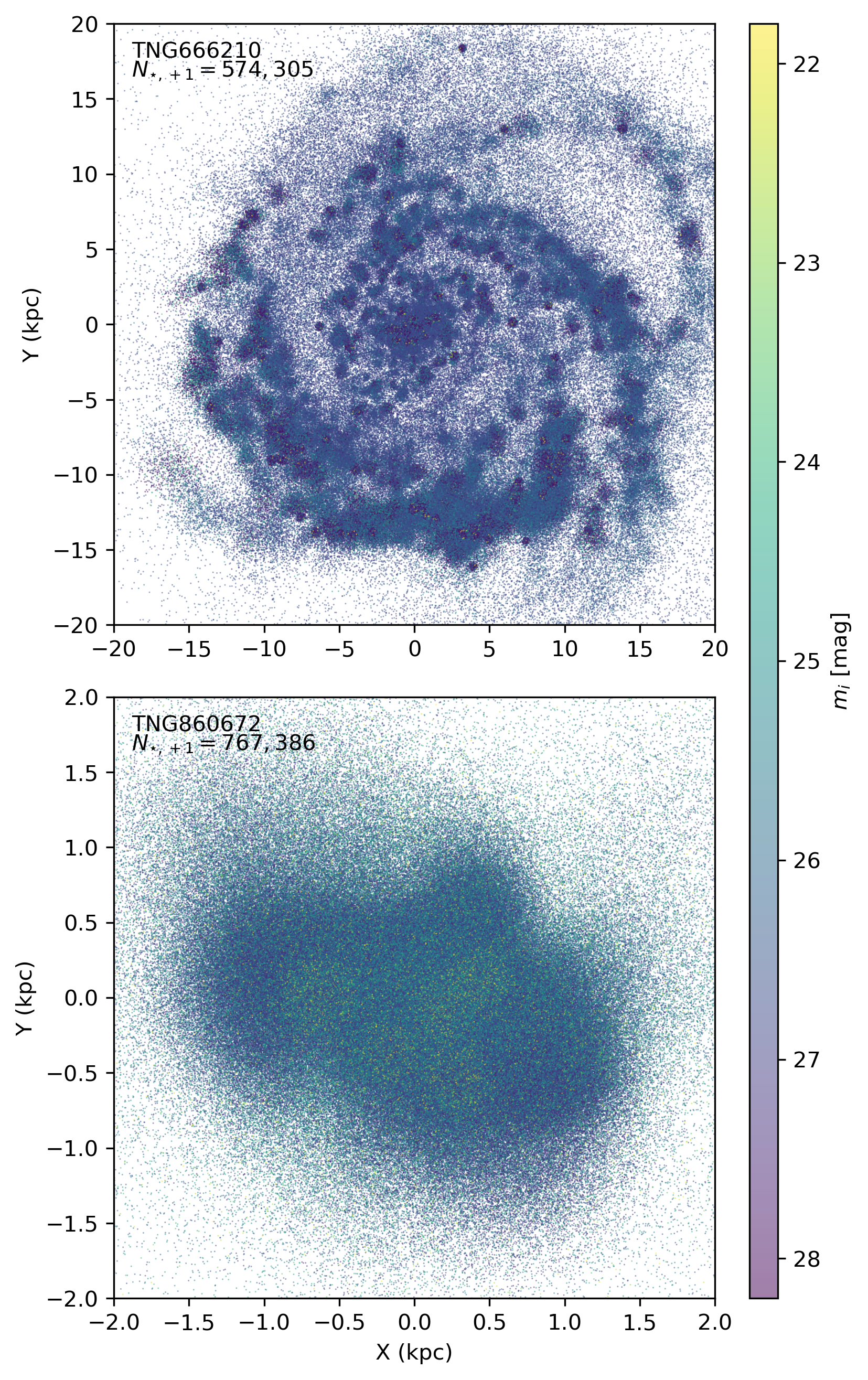}}
     \caption{Spatial distribution of resolved stars shown in a scatter plot in the CSST i band for TNG666210 (top) and TNG860672 (bottom). Colours indicate the CSST i-band apparent magnitudes of the stars. The annotated $N_{\star,\,+1}$ denotes the number of selected stars satisfying $m < m_\mathrm{lim} + 1$ in at least one of the seven CSST bands.}
     \label{fig2}
\end{figure}

With this cut, the retained resolved stars contribute about 49\% of the intrinsic total galaxy light in NUV, 31\% in u, and 16--20\% in the optical/NIR bands for TNG666210. For TNG860672, the resolved-to-total luminosity ratios are about 63\%--83\% over the u-to-y bands, while the NUV ratio exceeds unity by 4.8\%, owing to the oversubtraction discussed in Sect.~\ref{sec:dl}. The spatial distribution of these resolved stars is shown in Fig.~\ref{fig2}.

\subsection{Dust attenuation of resolved stars}\label{sec:attenuation}
\subsubsection{Diffuse ISM extinction}\label{sec:los}
All stars are subject to extinction by dust in the diffuse ISM. To estimate the line-of-sight extinction towards each star, we use the dust distribution reconstructed by SKIRT (Sect.~\ref{sec:skirt}) on an adaptive octree grid.

For each octree grid cell, SKIRT provides the cell volume $V$ and the monochromatic extinction coefficient $k_{\lambda}$ (per unit length). In this work, for each photometric band $b$, we evaluate $k_\lambda$ at the band pivot wavelength and denote it by $k_b$. For the CSST NUV, u, g, r, i, z, and y bands, the adopted pivot wavelengths are 0.286, 0.359, 0.471, 0.618, 0.760, 0.905, 0.964 $\mathrm{\mu m}$, respectively. We then approximate the optical depth of a cell as
\begin{equation}
\label{eq9}
\tau_{\mathrm{cell},\,b} = k_{b}\,l,
\end{equation}
where $l \equiv V^{1/3}$ is the side length of the cell. A more refined bandpass-dependent treatment will be explored in future work.

\begin{figure}
\centering
\includegraphics[width=\hsize]{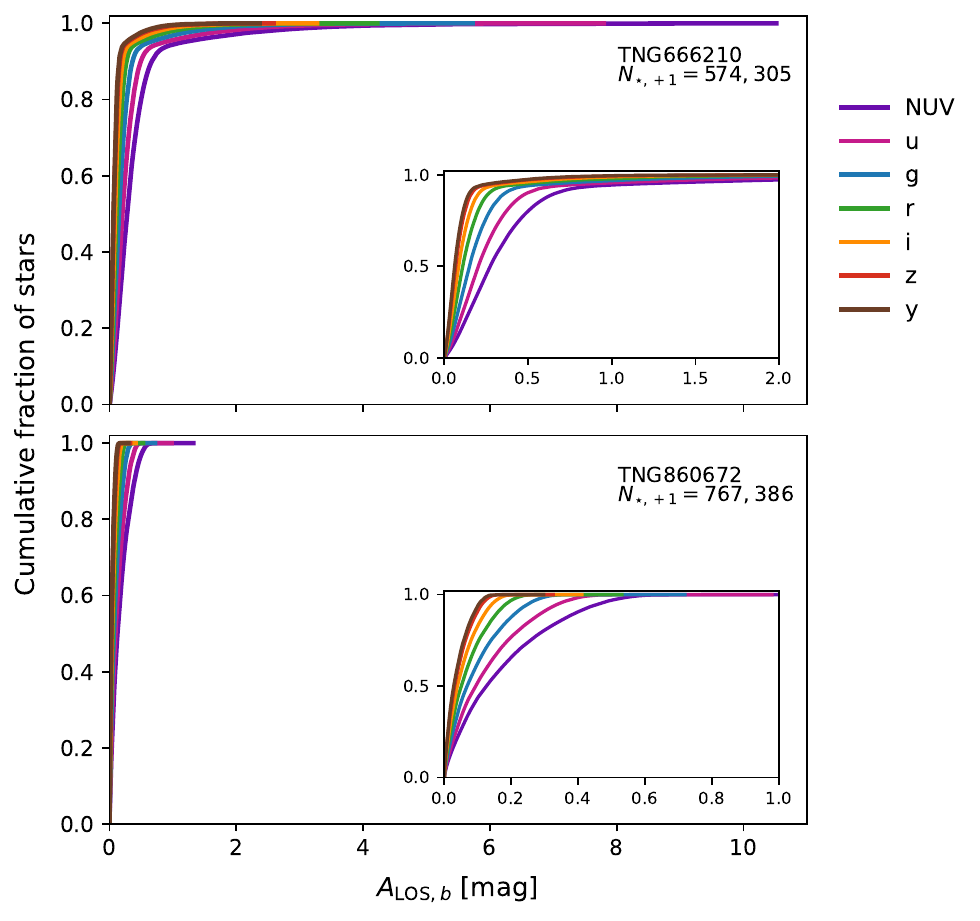}
\caption{Number-weighted CDFs of $A_{\mathrm{LOS},\,b}$ for resolved
stars in TNG666210 (top) and TNG860672 (bottom). The annotated $N_{\star,\,+1}$ has the same meaning as in Fig.~\ref{fig2}. Both main panels use the same $x$-axis range to include the high-extinction tail in TNG666210, while the insets zoom in on the low-extinction regime where most stars are located.}
\label{fig3}
\end{figure}

To compute the extinction towards a given star, both the stellar positions and the centres of the octree grid cells are transformed using the viewing geometry defined in Eq.~(\ref{eq7}) and projected onto the image plane. Each cell is then represented in projection as a square of side length $l$, centred on its projected position. For each star, we identify the foreground dust cells whose projected footprints contain the projected stellar position, and sum their optical depths along the line of sight. The total diffuse-ISM optical depth in each band is thus
\begin{equation}
\label{eq10}
\tau_{\mathrm{ISM},\,b} = \Sigma \tau_{\mathrm{cell},\,b},
\end{equation}
where the sum runs over all foreground cells satisfying the projected-overlap condition. The corresponding line-of-sight extinction is 
\begin{equation}
\label{eq11}
A_{\mathrm{LOS},\,b}=1.086\,\tau_{\mathrm{ISM},\,b}.
\end{equation}
Fig.~\ref{fig3} shows the resulting number-weighted cumulative distribution functions (CDFs) of $A_{\mathrm{LOS},\,b}$. To quantify both the typical extinction and the high-extinction tail, we also report the median, 90th-percentile, and 95th-percentile values of $A_{\mathrm{LOS},\,b}$ in Table~\ref{table:3}.

\begin{table}
\caption{Percentiles of $A_{\mathrm{LOS},\,b}$ for resolved stars.}
\label{table:3}
\centering
\begin{tabular}{lcccc}
\hline
Galaxy & Band & $A_{50}$ & $A_{90}$ & $A_{95}$ \\
       &      & [mag]    & [mag]    & [mag]    \\
\hline
TNG666210 & NUV & 0.274 & 0.667 & 1.143 \\
          & u & 0.203 & 0.494 & 0.847 \\
          & g & 0.149 & 0.363 & 0.622 \\
          & r & 0.110 & 0.268 & 0.459 \\
          & i & 0.085 & 0.207 & 0.356 \\
          & z & 0.068 & 0.164 & 0.281 \\
          & y & 0.062 & 0.150 & 0.257 \\
\hline
TNG860672 & NUV & 0.125 & 0.393 & 0.461 \\
          & u & 0.093 & 0.291 & 0.341 \\
          & g & 0.068 & 0.214 & 0.251 \\
          & r & 0.050 & 0.158 & 0.185 \\
          & i & 0.039 & 0.122 & 0.143 \\
          & z & 0.031 & 0.097 & 0.113 \\
          & y & 0.028 & 0.088 & 0.104 \\
\hline
\end{tabular}
\tablefoot{$A_{50}$, $A_{90}$, and $A_{95}$ are the 50th, 90th, and 95th percentiles of the corresponding extinction distribution.}
\end{table}

\subsubsection{Birth-cloud attenuation}
In addition to the line-of-sight extinction by the diffuse ISM, young stars in star-forming regions may experience additional attenuation by their birth clouds. For star particles younger than 10 Myr, we adopt the MAPPINGS III SED templates (Sect.~\ref{sec:radiation}), which self-consistently combine three components: the stellar-cluster spectrum generated with Starburst99 (2006 version; \citealt{1999ApJS..123....3L}), nebular emission from the associated H\,{\sc ii} region, and attenuation by the surrounding dusty PDR.

\begin{figure}
\centering
\includegraphics[width=0.85\hsize]{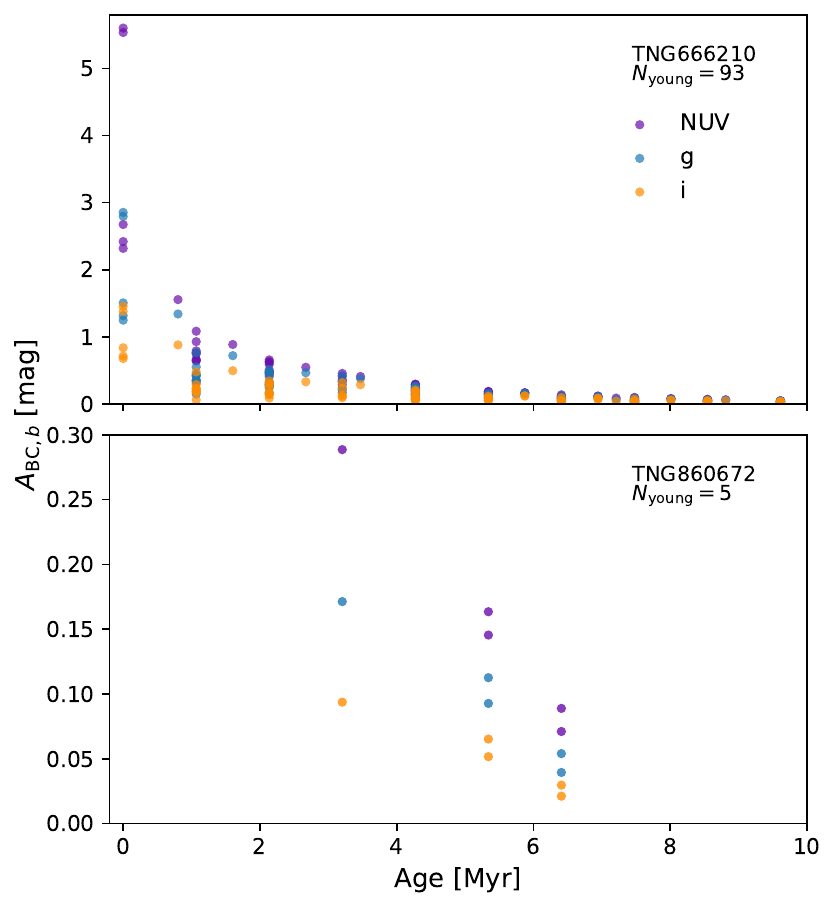}
\caption{Birth-cloud attenuation $A_{\mathrm{BC},\,b}$ as a function of young star particle age in TNG666210 (top) and TNG860672 (bottom). 
The attenuation is strongest at young ages and decreases rapidly 
with age, with larger attenuation at shorter wavelengths. 
At fixed age, the scatter in $A_{\mathrm{BC},\,b}$ reflects the dependence 
of the MAPPINGS III SEDs on additional particle properties, such as 
the SFR, metallicity, and H\,{\sc ii}-region compactness, 
besides the age-dependent PDR covering factor.}
\label{fig4}
\end{figure}

For each young particle, we compute the PDR covering fraction $f_{\mathrm {PDR}}$ as a function of age (Eq.~(\ref{eq2})). We interpret $f_{\mathrm {PDR}}=0$ as the limiting case of no surrounding PDR, i.e. no additional birth-cloud attenuation. For each CSST band, we then define the birth-cloud attenuation as
\begin{equation}
\label{eq12}
A_{\mathrm{BC},\,b} = -2.5\log_{10}
\left[
\frac{L_b(f_{\mathrm {PDR}})}{L_b(0)}
\right],
\end{equation}
where $L_b(f_{\mathrm {PDR}})$ and $L_b(0)$ are the band
luminosities computed with the age-dependent $f_{\mathrm {PDR}}$
and with $f_{\mathrm {PDR}}=0$, respectively. Fig.~\ref{fig4} shows $A_{\mathrm{BC},\,b}$ as a function of young star particle age. Here, $A_{\mathrm{BC},\,b}$ should be interpreted as an effective, particle-level attenuation derived from a composite MAPPINGS III SED, rather than as the exact attenuation for individual stars. We apply this attenuation to the resolved stars associated with the same parent young particle. This treatment therefore assumes a common mean birth-cloud attenuation for stars associated with the same star-forming region and does not capture star-to-star variations in local dust geometry or individual stellar SEDs.

\subsection{Unresolved diffuse light and image-level attenuation}\label{sec:dl}
\subsubsection{Diffuse-light residual image}
In addition to the resolved stars, the unresolved diffuse light, including faint stellar light and nebular emission from star-forming regions, contributes to the background that reduces the contrast for stellar detection. This effect becomes increasingly important at larger distances, where a greater fraction of the stellar light remains unresolved. 

To construct the unresolved diffuse-light images (Figs.~\ref{fig5}(c) and \ref{fig6}(c)), we use the transparent, noise-free probe maps output by SKIRT for each CSST band, as described in Sect.~\ref{sec:skirt_config}. These maps represent the emission of the imported star particles, computed from their assigned stellar population SEDs, distributed using the 3D SPH kernel defined in Eq.~(\ref{eq4}), and projected onto the image plane without photon-packet propagation. We subtract the resolved-star contribution from these maps as follows. For each star particle, we sum the fluxes of all associated resolved stars in the retained catalogue (Sect.~\ref{sec:catalogue}), and assign the total flux to the position of the parent particle. We then smooth this total flux with the
3D SPH kernel, and project the resulting flux onto the image plane. Finally, we subtract this reconstructed resolved-star image from the corresponding probe map to obtain the unresolved diffuse-light image. Specifically, for each band $b$, 
\begin{equation}
\begin{aligned}
I_{b,\,\mathrm{unres}}
 &= I_b-I_{b,\,\mathrm{res}} \\
 &= \left[
 I_{b,\,\mathrm{young}}
 - I_{b,\,\mathrm{young,\,res}}\,10^{-0.4A_{\mathrm{BC},\,b}}
 \right] \\
 &\quad+
 \left[
 I_{b,\,\mathrm{evolved}}
 - I_{b,\,\mathrm{evolved,\,res}}
 \right].
\end{aligned}
\label{eq:13}
\end{equation}
Here, $I_b$ is the intrinsic surface brightness image from the transparent, noise-free SKIRT probe map, and $I_{b,\,\mathrm{res}}$ is the reconstructed resolved-star image described above. The term $I_{b,\,\mathrm{young}}$ denotes the contribution from particles younger than 10 Myr, generated from the MAPPINGS III SEDs, while $I_{b,\,\mathrm{young,\,res}}$ is the reconstructed image of their associated resolved stars. The latter is attenuated by the birth-cloud attenuation $A_{\mathrm{BC},\,b}$ before subtraction. The first bracket in Eq.~(\ref{eq:13}) therefore represents the residual emission associated with the young particles, including unresolved stellar and nebular emission. The evolved-particle terms are
defined analogously.

In practice, the flux in the resolved-star image can locally exceed that in the corresponding intrinsic total-light image. This arises from differences in the stellar population models and IMFs used for the smooth stellar-particle light and the resolved-star catalogue. Specifically, the MAPPINGS III templates are based on Starburst99 SSP SEDs with a \citet{2001MNRAS.322..231K} IMF; the BC03 SSPs use Padova stellar evolutionary tracks with a \citet{2003PASP..115..763C} IMF; and the resolved stars are generated with TRILEGAL using PARSEC v1.2S stellar models and a \citet{2003PASP..115..763C} IMF. Moreover, the smooth stellar-particle luminosities are computed from IMF-averaged SEDs, whereas the TRILEGAL catalogue is generated through stochastic sampling. Local oversubtraction occurs mainly in regions dominated by relatively young stellar populations, where a small number of luminous massive stars can contribute a large fraction of the total light. As a result, individual TRILEGAL realisations may yield negative residual pixels.

To mitigate these artificial negative residuals, we replace the direct subtraction in Eq.~(\ref{eq:13}) with an age-separated, non-negative reconstruction when constructing the unresolved diffuse-light image. Specifically, star particles are divided into recent ($\leq 100\,\mathrm{Myr}$) and older ($>100\,\mathrm{Myr}$) components, as mentioned in Sect.~\ref{sec:skirt_config}. The 100 Myr threshold is chosen empirically based on an inspection of the particles responsible for the over-subtracted pixels, which are mostly younger than a few tens of Myr. It therefore provides a conservative boundary that separates the young populations, whose broad-band luminosities are most affected by stochastic sampling of luminous stars, from the older, more smoothly sampled component. This split is used only as a numerical device in the residual construction and does not modify the SED assignment described in Sect.~\ref{sec:radiation}. For each band, we compute the unresolved light separately for the recent and older components and set negative residual pixels to zero before summing:
\begin{equation}
\label{eq:14}
\begin{aligned}
I_{b,\,\mathrm{unres}}
 &= \max\left(I_{b,\,\mathrm{recent}}
 - I_{b,\,\mathrm{recent,\,res}},\,0\right) \\
 &\quad + \max\left(I_{b,\,\mathrm{older}}
 - I_{b,\,\mathrm{older,\,res}},\,0\right).
\end{aligned}
\end{equation}
This procedure is intended to provide a non-negative diffuse-light component for testing the stellar detection pipeline, rather than to enforce exact pixel-level flux conservation between the total-light and resolved-star images. Table~\ref{table:4} quantifies the resulting flux budget. For TNG666210, clipping negative residual pixels increases the reconstructed flux by 1.38\% in NUV and by less than 0.4\% in the other bands. For TNG860672, the effect is strongly concentrated in NUV: negative residuals occur in 16.7\% of the image pixels, while the corresponding regions contain about 65\% of the total intrinsic NUV SKIRT flux, indicating that the negative residuals are preferentially associated with compact NUV-bright regions. Clipping these residuals increases the total reconstructed flux by 8.14\%. The corresponding flux increase is below 0.5\% in all bands from u through y, where the residual amplitudes are substantially smaller than in NUV. Overall, clipping changes the reconstructed flux by less than 1.4\% in all cases except the NUV band of TNG860672.
\begin{table*}
\caption {Flux-budget check for clipping negative residual pixels in the construction of the unresolved diffuse-light image.}
\label{table:4} 
\centering
\tabcolsep 10pt
\begin{tabular}{ccccccccc}
\hline\hline 
Galaxy & Band & $F_\mathrm{int}$ & $F_\star$ & $F_\mathrm{raw,\,res}$ & $\Delta F_\mathrm{clip}$ & $F_\mathrm{final}$ & $\Delta F_\mathrm{clip}/F_\mathrm{int}(\%)$ & $f_\mathrm{neg}(\%)$\\ 
\hline
TNG666210 & NUV & 9.59e-25 & 4.71e-25 & 4.87e-25 & 1.32e-26 & 9.72e-25 & 1.38 & 2.8 \\
          & u & 1.39e-24 & 4.37e-25 & 9.57e-25 & 9.57e-28 & 1.40e-24 & 0.07 & 2.6 \\
          & g & 2.86e-24 & 5.35e-25 & 2.33e-24 & 1.09e-27 & 2.87e-24 & 0.04 & 3.0 \\
          & r & 3.96e-24 & 6.26e-25 & 3.33e-24 & 1.08e-27 & 3.96e-24 & 0.03 & 3.4 \\
          & i & 4.59e-24 & 8.35e-25 & 3.76e-24 & 4.34e-27 & 4.60e-24 & 0.09 & 8.3 \\
          & z & 5.47e-24 & 1.10e-24 & 4.37e-24 & 1.71e-26 & 5.49e-24 & 0.31 & 11.5 \\
          & y & 5.91e-24 & 1.20e-24 & 4.71e-24 & 2.36e-26 & 5.94e-24 & 0.40 & 12.2 \\
\hline
TNG860672 & NUV & 1.87e-24 & 1.96e-24 & -8.95e-26 & 1.52e-25 & 2.02e-24 & 8.14 & 16.7\\  
          & u & 2.38e-24 & 1.98e-24 & 4.02e-25 & 1.08e-26 & 2.39e-24 & 0.45 & 7.8 \\
          & g & 3.43e-24 & 2.73e-24 & 7.00e-25 & 8.08e-27 & 3.44e-24 & 0.24 & 6.5 \\
          & r & 4.35e-24 & 2.99e-24 & 1.37e-24 & 4.10e-27 & 4.36e-24 & 0.09 & 6.0 \\
          & i & 4.10e-24 & 3.22e-24 & 8.78e-25 & 2.66e-27 & 4.10e-24 & 0.06 & 6.4 \\
          & z & 4.85e-24 & 3.48e-24 & 1.38e-24 & 1.33e-27 & 4.85e-24 & 0.03 & 5.6 \\
          & y & 5.64e-24 & 3.57e-24 & 2.07e-24 & 7.65e-28 & 5.64e-24 & 0.01 & 5.2 \\
\hline
\end{tabular}
\tablefoot{$F_\mathrm{int}$ is the intrinsic total flux obtained by summing $I_b$ over the image, $F_\star$ is the flux of the explicit resolved-star component after applying the birth-cloud attenuation to stars associated with young particles, consistent with the definition of
$F_\mathrm{int}$, and $F_\mathrm{raw,\,res} = F_\mathrm{int}-F_\star$ is the raw residual before clipping negative pixels. $\Delta F_\mathrm{clip}$ is the additional flux introduced by setting negative residual pixels to zero, and $F_\mathrm{final}=F_\star+F_\mathrm{raw,\,res}+\Delta F_\mathrm{clip}$ is the final reconstructed flux. $f_\mathrm{neg}$ is the fraction of image pixels with negative raw residual values. All fluxes are in units of erg s$^{-1}$ cm$^{-2}$ Hz$^{-1}$.}
\end{table*}

\subsubsection{Image-level effective attenuation map}
To account for the dust effect on the unresolved diffuse-light image, we construct an image-level effective attenuation-factor map from the SKIRT simulations. For each band $b$, we first compute
\begin{equation}
\label{eq:15}
R_{b,\,\mathrm{eff}}(x,y)=
\frac{I_{b,\,\mathrm{dust}}(x,y)}{I_{b,\,\mathrm{trans}}(x,y)},
\end{equation}
where $I_{b,\,\mathrm{dust}}$ includes both the direct and scattered components, and $I_{b,\,\mathrm{trans}}$ is the corresponding transparent total-light RT image. The ratio is evaluated within pixels satisfying $I_{b,\,\mathrm{trans}}>10^{-3}I^\mathrm{max}_{b,\,\mathrm{trans}}$. We then define the masked attenuation-factor map $R_{b,\,\mathrm{eff,\,mask}}$ by retaining these values within the mask and setting $R_{b,\,\mathrm{eff,\,mask}}=1$ outside it. No additional smoothing or percentile clipping is applied. All finite positive $R_{b,\,\mathrm{eff}}$ values within the mask are retained. For reference, the maximum retained $R_{b,\,\mathrm{eff}}$ values span 1.27--2.89 for TNG666210 and 1.06--2.33 for TNG860672 across the seven bands, corresponding to the most negative attenuation values of -1.15 to -0.26 mag and -0.92 to -0.06 mag, respectively.

\begin{figure*}
    \centering
    {\includegraphics[width=0.75\textwidth]{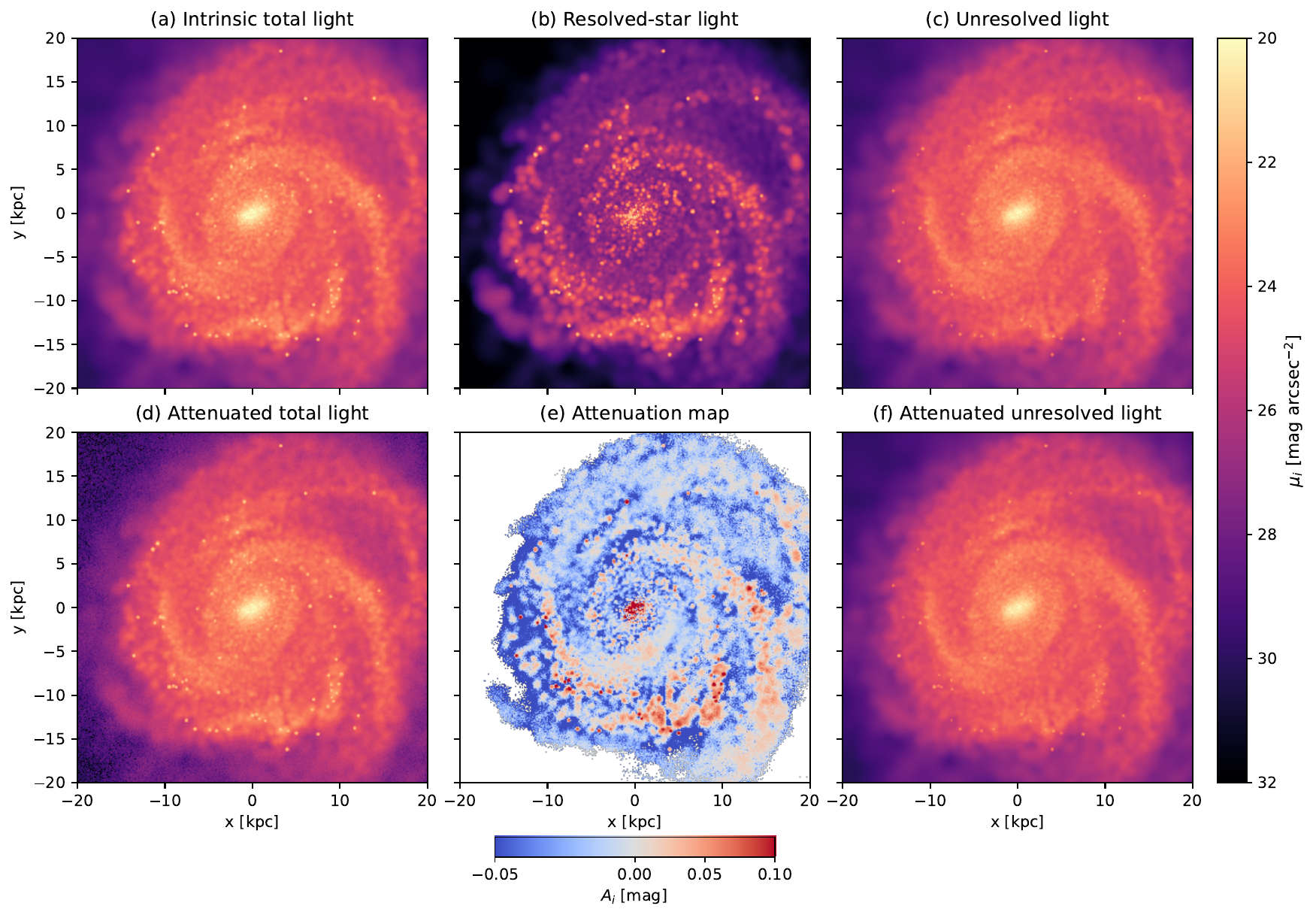}}
     \caption{Illustration of the construction of the dust-attenuated unresolved diffuse light in the CSST i band for TNG666210. Panels (a) and (b) show the intrinsic total surface brightness from the transparent, noise-free SKIRT probe map and the resolved-star contribution, respectively, with the latter reconstructed by assigning the fluxes of resolved stars in the retained catalogue to their parent particles, smoothing this total flux with the 3D SPH kernel, and projecting it onto the image plane; their difference gives the unresolved diffuse light in panel (c). Panel (d) shows the dust-attenuated total-light RT image produced by SKIRT. Panel (e) shows the image-level effective attenuation map, expressed as $A_{i,\,\mathrm{eff}} = -2.5\log_{10} R_{i,\,\mathrm{eff,\,mask}}$, where $R_{i,\,\mathrm{eff,\,mask}}$ is the masked attenuation-factor map. For visualisation, pixels outside the adopted $10^{-3}I^{\max}_{i,\,\mathrm{trans}}$ mask are shown in white; in the actual application to panel (c), these pixels are assigned $R_{i,\,\mathrm{eff,\,mask}}=1$. Panel (f) shows the final dust-attenuated unresolved diffuse-light image obtained by multiplying panel (c) by $R_{i,\,\mathrm{eff,\,mask}}$.}
      \label{fig5}
\end{figure*}

\begin{figure*}
    \centering
    {\includegraphics[width=0.75\textwidth]{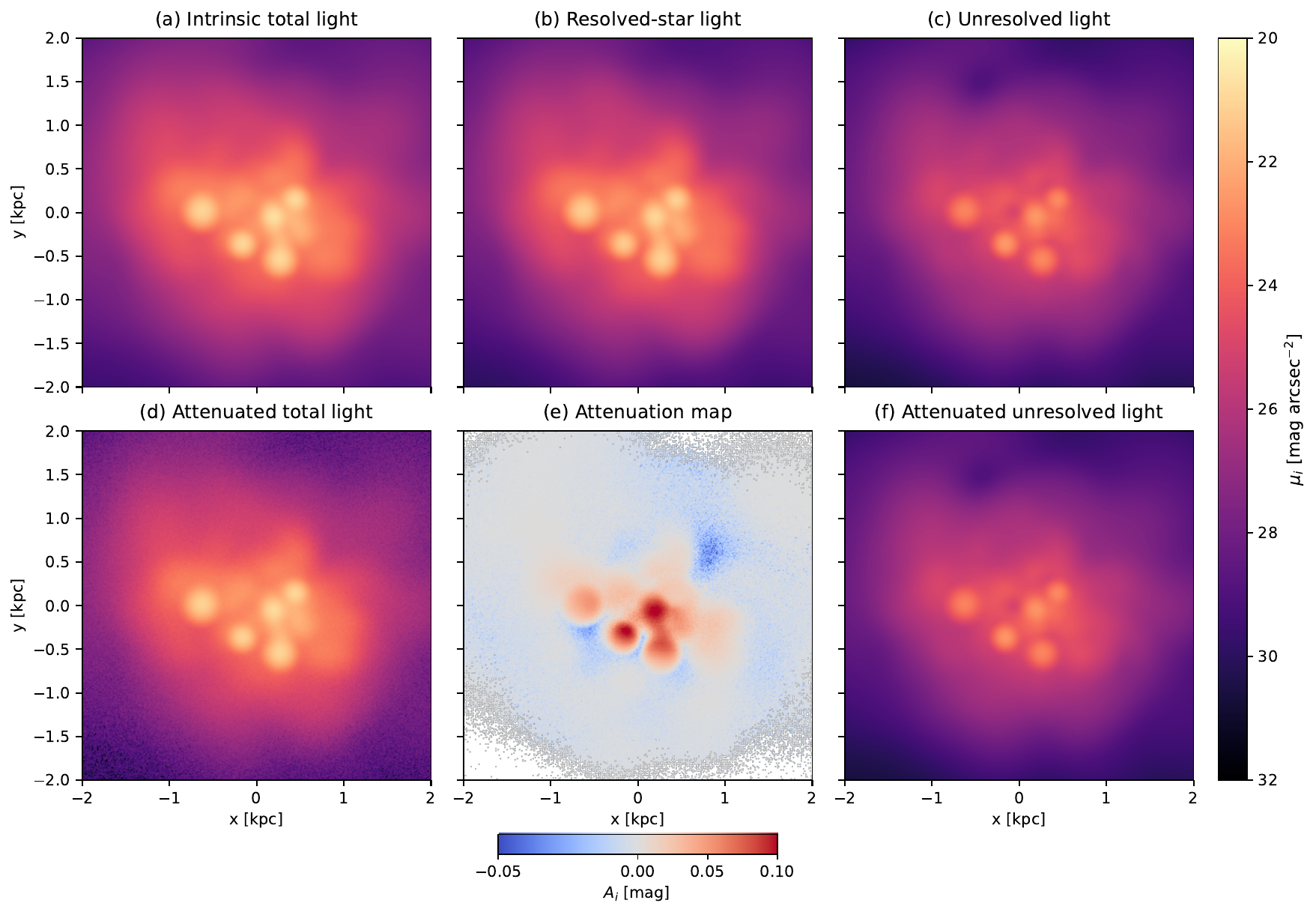}}
     \caption{Same as Fig. \ref{fig5}, but for TNG860672.}
      \label{fig6}
\end{figure*}

The processed effective attenuation map is then defined as
\begin{equation}
\label{eq:16}
A_{b,\,\mathrm{eff}}(x,y)=-2.5\log_{10}R_{b,\,\mathrm{eff,\,mask}}(x,y),
\end{equation}
The negative values in attenuation maps (Figs.~\ref{fig5}(e) and \ref{fig6}(e)) arise because $I_{b,\,\mathrm{dust}}$ includes scattered photons, and $R_{b,\,\mathrm{eff,\,mask}}$ can exceed unity when scattered light entering the line of sight overcompensates the loss of direct light relative to the transparent image. To assess whether this excess is robust against MC noise, we perform six independent SKIRT realisations and compare $I_{b,\,\mathrm{dust}}-I_{b,\,\mathrm{trans}}$ with the run-to-run scatter. We note that the measured scatter also includes small variations introduced by the stochastic octree dust-grid sampling. For TNG666210, about half to two-thirds of all negative pixels exceed the 2$\sigma$ run-to-run scatter. This fraction increases to about 75--95\% for the stronger half of the negative pixels selected by $|A_{b,\,\mathrm{eff}}|$ in each band. For TNG860672, the fraction of statistically significant negative pixels is high in the blue bands but drops towards the red bands, where the median negative attenuation is only a few millimagnitudes. We therefore interpret the coherent and stronger negative structures as scattering-dominated, while weak near-zero negative pixels are more affected by stochastic variations in the RT calculation.

The dust-attenuated unresolved diffuse-light image is obtained as
\begin{equation}
\label{eq:17}
I^\mathrm{dust}_{b,\,\mathrm{unres}}(x,y)
=
I_{b,\,\mathrm{unres}}(x,y)R_{b,\,\mathrm{eff,\,mask}}(x,y).
\end{equation}

As a check on the attenuation mask choice, we compare the dust-attenuated unresolved diffuse-light images obtained with the fiducial $10^{-3}I^\mathrm{max}_{b,\,\mathrm{trans}}$ mask and a looser $10^{-4}I^\mathrm{max}_{b,\,\mathrm{trans}}$ mask. For TNG666210, the
total diffuse flux differs by 2.1\% in NUV, 0.6\% in u, and less than
0.2\% in the remaining bands. For TNG860672, the corresponding differences are 1.4\%, 0.1\%, and less than 0.02\%. Thus the mask choice has only a small effect on the final dust-attenuated unresolved diffuse-light images.

Finally, we emphasise that $R_{b,\,\mathrm{eff,\,mask}}$ is used only for the unresolved diffuse-light image and is not assumed to be equivalent to the line-of-sight extinction $A_{\mathrm{LOS},\,b}$ applied to resolved stars in Sect.~\ref{sec:los}. As a check, we sample the corresponding $A_{b,\,\mathrm{eff}}$ at the positions of the resolved stars and use these values in place of $A_{\mathrm{LOS},\,b}$ solely to quantify the impact of the different dust treatments. This substitution would increase the integrated resolved-star flux by 7.6--14.6\% for TNG666210 and 2.1--7.6\% for TNG860672 across the bands. The median values of the per-star magnitude difference, $m_\mathrm{map}-m_\mathrm{LOS}$, range from -0.15 mag in NUV to -0.06 mag in y for TNG666210, and from -0.13 to -0.02 mag for TNG860672. These treatments are not expected to yield identical results: the image-level attenuation map is derived from bandpass-integrated SKIRT images and therefore includes contributions from sources at different depths, together with absorption and scattering through the 3D dust distribution. In contrast, the extinction for each resolved star is estimated from the dust optical depth between that star and the observer at the pivot wavelength of each band.

\section{Catalogue-level effects of dust attenuation on resolved stellar populations}\label{sec:dust_effect} 
For the two TNG50 galaxies considered here, we quantify the impact of dust attenuation, including diffuse ISM extinction and birth-cloud attenuation, by applying the nominal CSST 5$\sigma$ point-source limiting magnitudes to both the intrinsic and attenuated stellar catalogues. Specifically, we select stars satisfying $m < m_\mathrm{lim}$ in at least one of the seven CSST bands, and refer to them as ``selected stars'' throughout this section. The resulting counts are catalogue-level estimates rather than image-level source counts obtained through source detection and recovery in mock images.

Table~\ref{table:5} summarises the resulting changes in the catalogue-level counts. Dust attenuation reduces the number of selected stars in both galaxies, with a stronger effect in TNG666210 than in TNG860672. The
reduction is strongly wavelength dependent, being largest in NUV
and much weaker in the i band, consistent with stronger dust
attenuation at shorter wavelengths.
\begin{table}
\caption{Effect of dust attenuation on the catalogue-level counts of selected stars.}
\label{table:5}
\centering
\begin{tabular}{lcrrrr}
\hline
Galaxy & Band & $N_{\rm int}$ & $N_{\rm att}$ & $\Delta N$ & $\Delta N/N_{\rm int}$ \\
\hline
TNG666210 & Any & 148498 & 117977 & 30521 & 20.55\% \\
          & NUV & 39067  & 21665  & 17402 & 44.54\% \\
          & i & 94514  & 80287  & 14227 & 15.05\% \\
\hline
TNG860672 & Any & 412644 & 391601 & 21043 & 5.10\% \\
          & NUV & 35915  & 30405  & 5510  & 15.34\% \\
          & i & 285410 & 276993 & 8417  & 2.95\% \\
\hline
\end{tabular}
\tablefoot{$N_{\rm int}$ and $N_{\rm att}$ denote the numbers of selected stars before and after attenuation, respectively, with
$\Delta N = N_{\rm int}-N_{\rm att}$. ``Any'' denotes stars satisfying $m < m_\mathrm{lim}$ in at least one of the seven CSST bands.}
\end{table}

\begin{figure}
    \centering
    {\includegraphics[width=0.85\hsize]{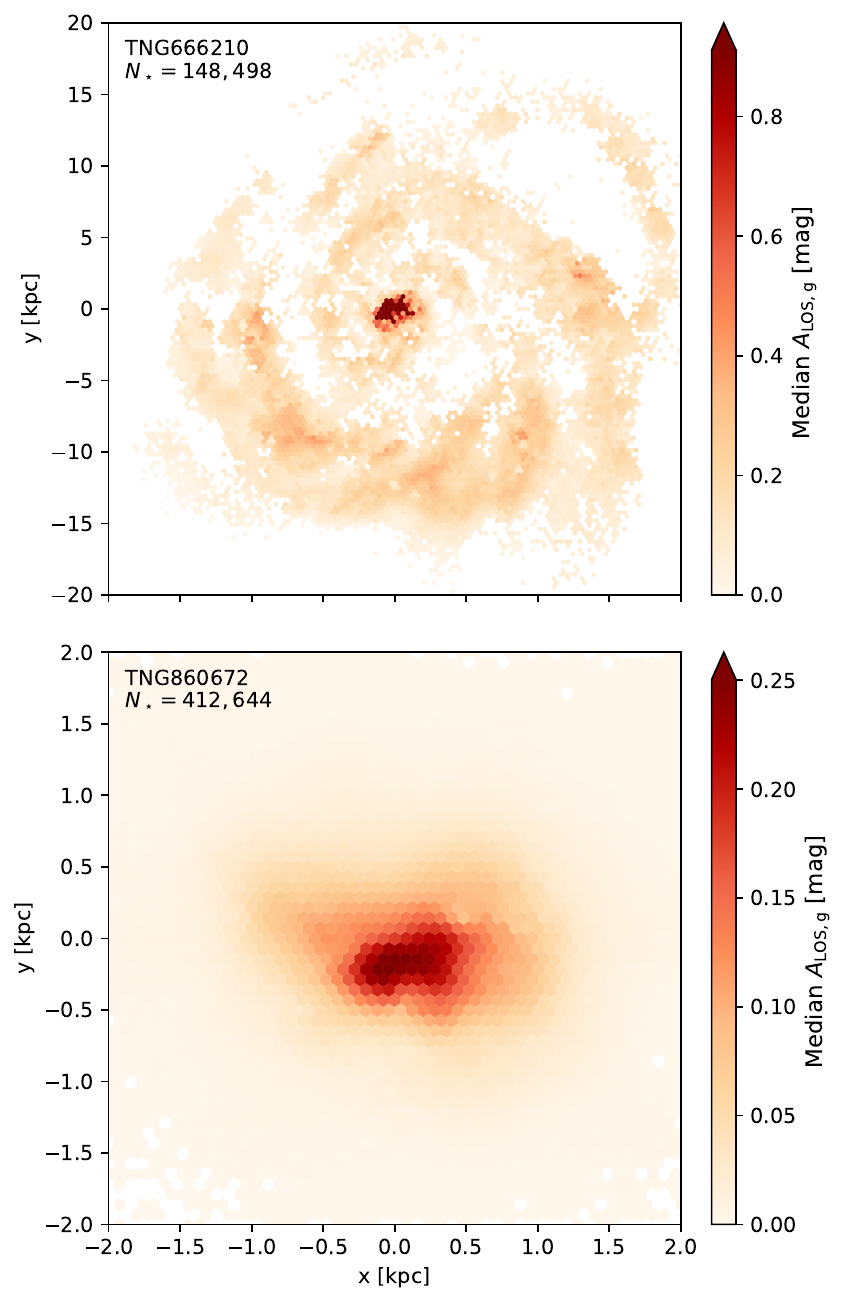}}
     \caption{Hexagonal-binned spatial distribution of the median line-of-sight extinction in the CSST g band, $A_{\mathrm{LOS},\,g}$, for TNG666210 (top) and TNG860672 (bottom). $N_\star$ denotes the number of selected stars satisfying $m < m_\mathrm{lim}$ in at least one of the seven CSST bands. Colours indicate the median $A_{\mathrm{LOS},\,g}$ in each spatial bin. For each galaxy, the colour scale is truncated at the 95th percentile of the positive $A_{\mathrm{LOS},\,g}$ values to enhance spatial contrast.}
     \label{fig7}
\end{figure}

The spatial extinction maps in Fig.~\ref{fig7} provide a complementary view of this effect. In TNG666210, the median $A_{\rm LOS,\,g}$ is strongly
structured, with enhanced extinction in the central region and along
the spiral arms, leading to a more pronounced reduction in the number of selected stars. By contrast, TNG860672 exhibits a centrally concentrated extinction distribution with lower overall amplitudes, consistent with its smaller count reduction. We note that these extinction maps trace only the line-of-sight diffuse-ISM extinction, whereas the attenuated stellar magnitudes used in the colour--magnitude diagrams (CMDs) below additionally include the birth-cloud attenuation for young stars.

\begin{figure*}
\sidecaption
  \includegraphics[width=11.8cm]{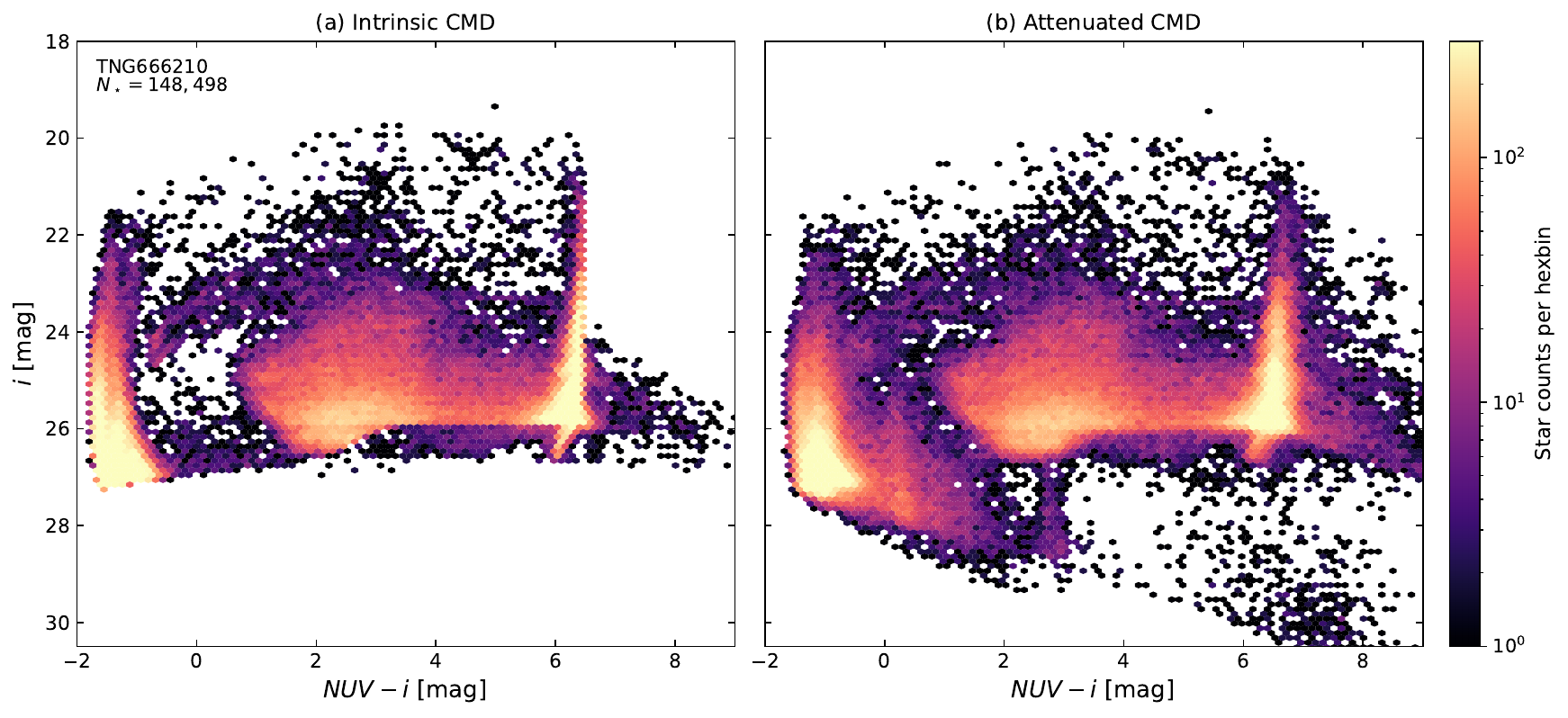}
     \caption{Hexagonal-binned CMDs of TNG666210 in the NUV--i versus i plane, shown before (left) and after (right) dust attenuation. The attenuation includes contributions from both the diffuse ISM and birth clouds. The attenuated CMD is shifted towards redder colours and fainter magnitudes.}
     \label{fig8}
\end{figure*}

\begin{figure*}
\sidecaption
  \includegraphics[width=11.8cm]{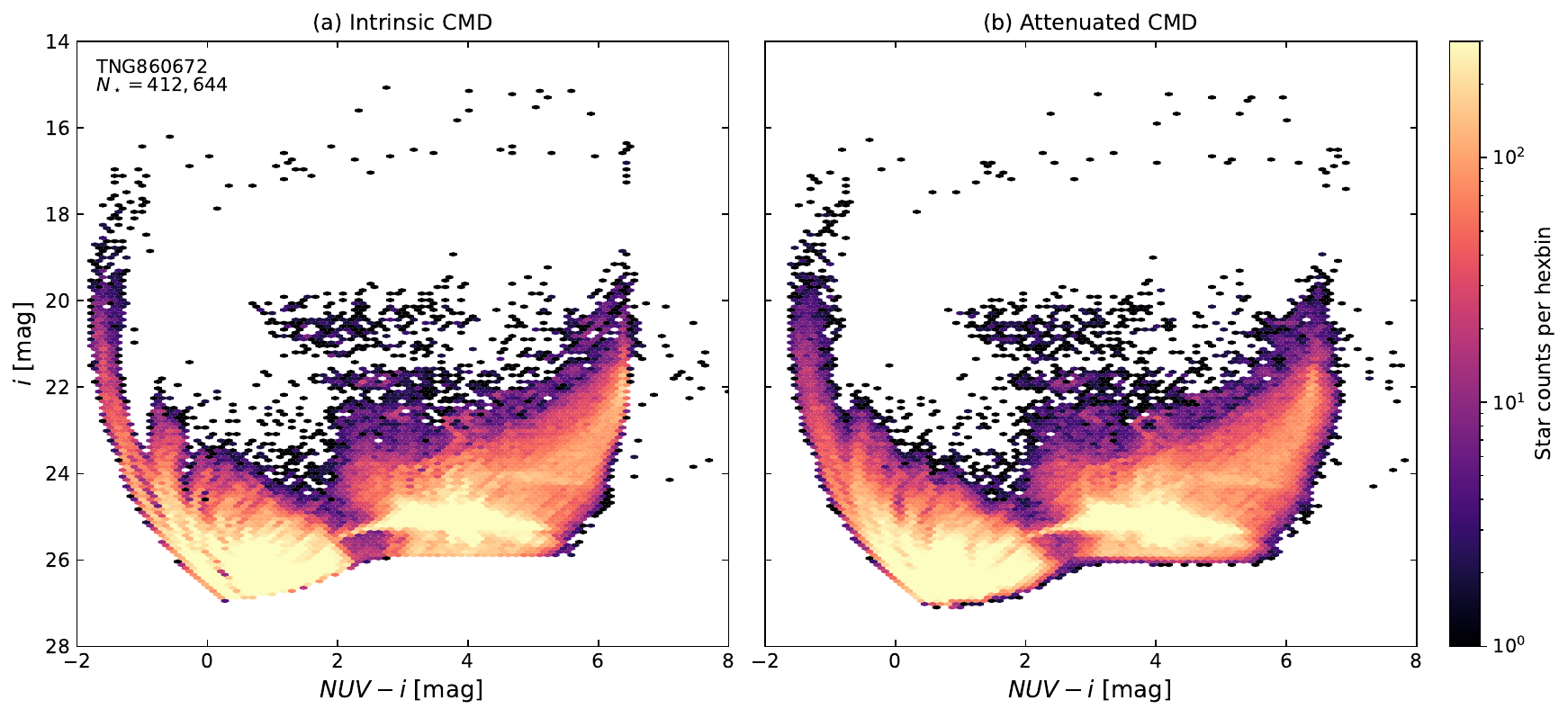}
     \caption{Same as Fig.~\ref{fig8}, but for TNG860672. The gap at NUV--i $\sim$ 1--6 and i $\sim$ 18--20 reflects the absence of stellar particles with ages between 10 and 40 Myr.}
     \label{fig9}
\end{figure*}

The CMDs further illustrate how dust attenuation modifies the resolved stellar population. Figs.~\ref{fig8} and \ref{fig9} show that dust attenuation shifts stars towards redder NUV--i colours and fainter i-band magnitudes, reflecting the stronger attenuation in the NUV than in the i band. It also broadens the blue main-sequence features, especially in TNG666210, where the stronger and more spatially structured extinction produces a more extended red side of the CMD. These effects are weaker in TNG860672, whose CMD morphology
is only mildly affected. The gap at NUV--i $\sim$ 1--6 and i $\sim$ 18--20 
in TNG860672 reflects the absence of stellar particles with ages between
10 and 40 Myr, rather than dust attenuation.

For the two examples, at the idealised catalogue level, dust attenuation reduces the number of selected stars and alters the morphology of the resulting CMDs. A full image-level assessment of these dust-attenuation effects, including source detection, photometric recovery, and completeness analyses in the presence of crowding and complex image backgrounds, will be addressed in future work. 

\section{Mock CSST images}\label{sec:mock}
\citet{2026RAA....26b4001W} provide a comprehensive, high-fidelity simulation pipeline for generating CSST imaging observations. This pipeline is built on the GalSim \citep{2015A&C....10..121R} package, incorporating the CSST instrumental specifications and lab-measured characteristics. In this work, we adopt the released version 3.3.0 of this pipeline\footnote{\url{https://csst-tb.bao.ac.cn/code/csst-sims/csst_msc_sim/-/tree/v3.3.0}} to generate mock CSST images from the attenuated stellar catalogues, together with the corresponding attenuated unresolved diffuse-light images. The diffuse-light images supplied to the pipeline have a pixel scale of $\sim 1.03''\,\mathrm{pixel^{-1}}$, and are resampled to the CSST pixel scale of $0.074''\,\mathrm{pixel^{-1}}$ using GalSim's \texttt{InterpolatedImage}.

We perform the simulations in the seven CSST photometric
bands, NUV, u, g, r, i, z, and y, and generate single-exposure science-level images including PSF convolution, sky background, stray light, photon Poisson noise, dark current and associated noise, and readout noise. Although the pipeline supports the inclusion of foreground and background contaminants, such as Milky Way stars and distant galaxies, these are omitted here to focus on the properties of the simulated galaxies’ resolved stellar populations. Multiple dithered exposures and their subsequent alignment, resampling and co-addition are not supported, and hence correlated resampling noise is not included. Cosmic rays and detector defects can be modelled but are not enabled in the present simulations. Photometric and astrometric calibration uncertainties are neglected. The photometric zero points and the world coordinate system (WCS) transformation between sky coordinates and detector pixels are treated as exact by construction. Field distortion is not enabled in the simulations. The images are digitised as 16-bit data and stored in digital numbers (DN), assuming a gain of $G = 1~e^-\,{\rm DN}^{-1}$. Pixel values are clipped at the digital saturation limit of 65{,}535 DN.  

For the resolved stellar point sources, we adopt spatially varying PSFs in each photometric band. The band-dependent PSF at each source position is constructed as a wavelength-weighted combination of four monochromatic PSFs sampled across the corresponding filter bandpass. The unresolved diffuse-light images are convolved with a band-dependent PSF evaluated at the field centre and constructed using the same wavelength-weighting procedure.

For the astrometric input, the projected stellar coordinates are
converted to RA and Dec using a TAN-projection WCS,  with the galaxy centre placed at an arbitrary sky position of RA = 150 deg and Dec = 30 deg. This choice only defines the sky-coordinate system and does not affect the intrinsic properties of the mock galaxies.

The final mock images, shown in Figs.~\ref{fig10} and \ref{fig11} for the NUV and i bands, are tailored to $8080 \times 8080$ pixels. At the CSST pixel scale of $0.074\,{\rm arcsec\,pixel^{-1}}$, this corresponds to $\sim$29 kpc × 29 kpc for TNG666210 and $\sim$2.9 kpc × 2.9 kpc for TNG860672. The input stellar catalogues contain 515{,}930 stars for TNG666210 and 735{,}525 stars for TNG860672 in the corresponding FoV. The runtime per $10^5$ input stars is typically $\sim$10 hours on a single CPU core. These mock images serve as the basis for image-level source detection, photometric recovery, and completeness analysis in subsequent studies. The datasets, including mock images, input stellar catalogues, astrometric/photometric calibration information, and supporting materials for reproducibility, are publicly available via the Chinese Virtual Observatory Data Center{\footnote{\url{https://nadc.china-vo.org/res/r101856/}}.
\begin{figure*}
    \centering
    {\includegraphics[width=\textwidth]{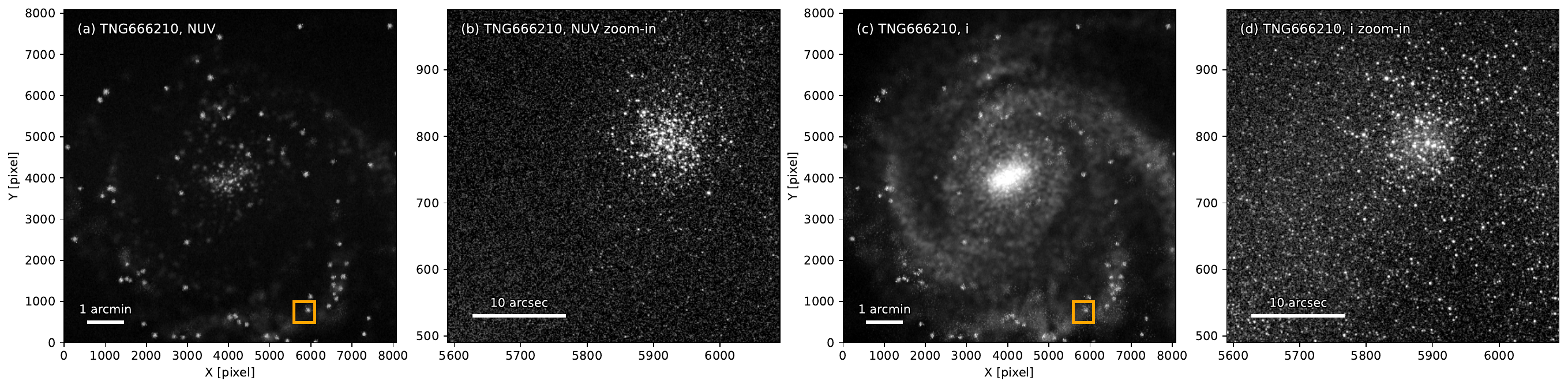}}
     \caption{Mock CSST images of TNG666210 in the NUV (a, b) and i (c, d)  bands. The exposure time is 150 s per band; pixel values are given in DN, assuming a gain of $G = 1~e^-\,{\rm DN}^{-1}$. Images are displayed with an asinh stretch, with the lower bound set to near median background level (4 DN for NUV and 35 DN for i). The panels (a, c) show the full galaxy, while the yellow boxes mark the regions shown in the zoom-in panels (b, d). The 10-arcsec scale bar corresponds to $\sim$485 pc at the adopted distance of 10 Mpc.}
      \label{fig10}
\end{figure*}

\begin{figure*}
    \centering
    {\includegraphics[width=\textwidth]{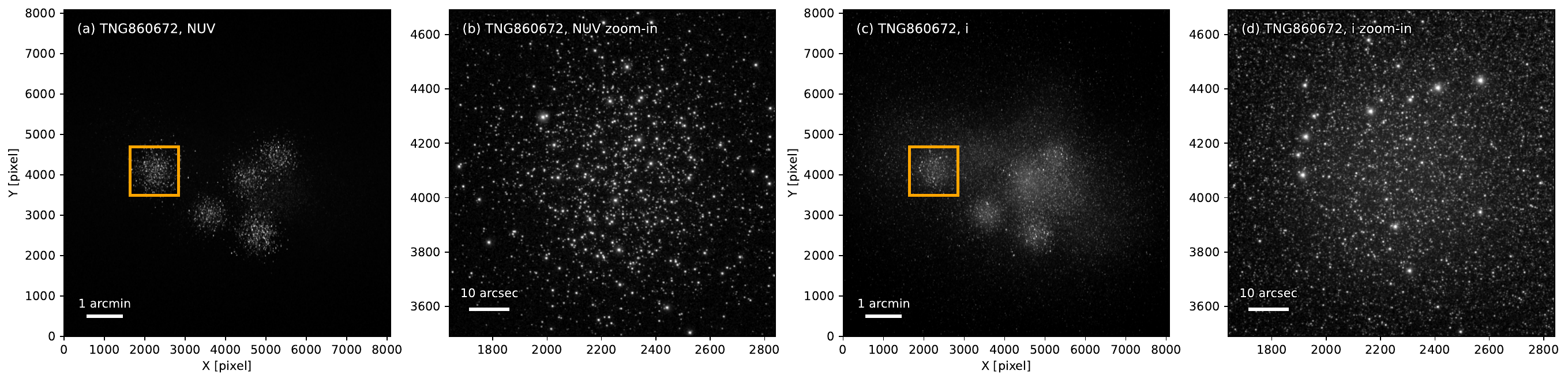}}
     \caption{Same observing setup and layout as Fig.~\ref{fig10}, but for TNG860672 at an adopted distance of 1 Mpc. The images are displayed with an independently adjusted asinh stretch, with lower bounds of 4 DN for NUV and 35 DN for i. The 10-arcsec scale bar in the zoom-in panels corresponds to $\sim$48.5 pc.}
      \label{fig11}
\end{figure*}

\section{Summary and outlook}\label{sec:summary}
The forthcoming CSST imaging surveys will provide a unique opportunity to study
resolved stellar populations in nearby galaxies with space-based resolution and
broad wavelength coverage. To fully exploit these data, realistic mock
observations are essential for forward-modelling simulated galaxies into
observable images, for assessing the recovery of stellar catalogues, and for quantifying observational effects such as crowding, dust attenuation, unresolved diffuse light, sky background, and instrumental noise. In this work,
we developed a framework for generating mock CSST images of TNG50 galaxies
designed for studies of resolved stellar populations. The framework links the
intrinsic stellar populations represented by the simulation star particles to dust-attenuated stellar catalogues, unresolved diffuse-light maps, and mock CSST images.

For each selected TNG50 galaxy, we generated stellar catalogues from TNG50 star particles using TRILEGAL, assigning intrinsic multi-band magnitudes to each synthetic star in the seven CSST photometric bands, NUV, u, g, r, i, z, and y.
Stellar positions were assigned separately for evolved and young particles,
using local anisotropic 3D Gaussian distributions with covariance matrices
estimated from neighbouring star particles for the former, and isotropic
Gaussian distributions with widths tied to the particle smoothing lengths for
the latter. We adopted a magnitude cut, defined relative to the
nominal 5$\sigma$ CSST point-source limiting depths, to separate stars represented
explicitly as point sources from the remaining unresolved stellar light. The
intrinsic unresolved diffuse-light images were constructed by subtracting the
kernel-smoothed reconstructed image of the sampled resolved stars from the
intrinsic total-light image of the star particles, with an age-based split used
to suppress artificial negative residuals.

We then modelled dust attenuation using a dust distribution inferred from the
gas mass and metallicity distributions in the TNG50 simulations. For the
resolved stars, line-of-sight extinction by the diffuse ISM was computed by
integrating the dust optical depth through the SKIRT adaptive octree grid, and
birth-cloud attenuation was additionally applied to young stars. Using SKIRT, we also computed dust-attenuated total-light images and transparent reference
images, from which we derived image-level transmission maps used to attenuate the unresolved diffuse-light images.

We applied this framework to two illustrative TNG50 galaxies with contrasting
morphologies, the face-on spiral TNG666210 and the dwarf irregular TNG860672.
For these two systems, we first examined at the ideal-catalogue level how dust
attenuation modifies the intrinsic stellar distributions in colour--magnitude
space and affects the visibility of resolved stellar populations. We then
passed the dust-attenuated resolved-star catalogues and unresolved diffuse-light
images to the CSST simulation pipeline. The final mock observations include PSF
convolution, sky background, stray light, photon Poisson noise, dark current
and associated noise, and readout noise. These examples illustrate how stellar structure, dust distribution, and unresolved diffuse
light shape the final CSST-like mock images in the two selected systems, which serve as illustrative test cases and do not constitute a systematic exploration of nearby-galaxy properties. Future work will explore the dependence of the mock images on galaxy properties, viewing angle, distance, and dust-related assumptions.

These mock images are designed to serve as a realistic image-level testbed for future validation of CSST resolved-star photometry pipelines. Together with their known input catalogues, they will enable controlled assessments of how crowding, dust attenuation, unresolved diffuse light, and instrumental effects affect source detection, photometric accuracy, completeness, and related pipeline performance. The resulting truth-matched data products can also support future studies of how these observational effects propagate into stellar-parameter inference and CMD-based SFH recovery. More broadly, because
the framework treats the astrophysical input catalogues, unresolved
diffuse-light modelling, dust attenuation treatment, and instrument-specific
image simulation as distinct components, it can be readily adapted to other hydrodynamical simulations and multi-band imaging facilities.

\begin{acknowledgements}
      We thank the anonymous referee for their helpful comments and suggestions. This work is based on the software created by the CSST Simulation Team, which is supported by the CSST scientific data processing and analysis system of the China Manned Space Project. This work is supported by the National Key R\&D Program of China No. 2025YFF0511001. This work is supported by Chinese Academy of Sciences President's International Fellowship Initiative grant No. 2026PVA0172, 2024PD0058. J.T. acknowledges the support by the China Manned Space Project with grant No. CMS-CSST-2025-A20, CMS-CSST-2025-A11. R.L. acknowledges the science research grants from the China Manned Space Program No. CMS-CSST-2025-A03, CAS Project for Young Scientists in Basic Research No. YSBR-062. Y.C. acknowledges the science research grants from the China Manned Space Program No. CMS-CSST-2025-A14, the National Natural Science Foundation of China (NSFC) No. 12573030, No. 12533008, and the Natural Science Research Project of Anhui Educational Committee No. 2024AH050049.
\end{acknowledgements}

\bibliographystyle{bibtex/aa}
\bibliography{bibtex/bibtex}

\begin{appendix}
\onecolumn
\section{Dust density reconstruction}
We compare the input and SKIRT-gridded dust density fields for TNG666210 and
TNG860672 using planar cuts through the three coordinate planes
(Figs.~\ref{figa1} and \ref{figa2}). Quantitatively, the total dust mass is conserved to better than 0.01\% for TNG666210 and differs by -0.08\% for
TNG860672. The optical depths at $0.55\,\mu{\rm m}$ along the full $x/y/z$
axes agree within 6\% for TNG666210 and within 4\% for TNG860672. These
results indicate that the adopted spatial grid adequately reconstructs the
input dust density fields.
\begin{figure*}[ht!]
    \centering
    {\includegraphics[width=0.8\textwidth]{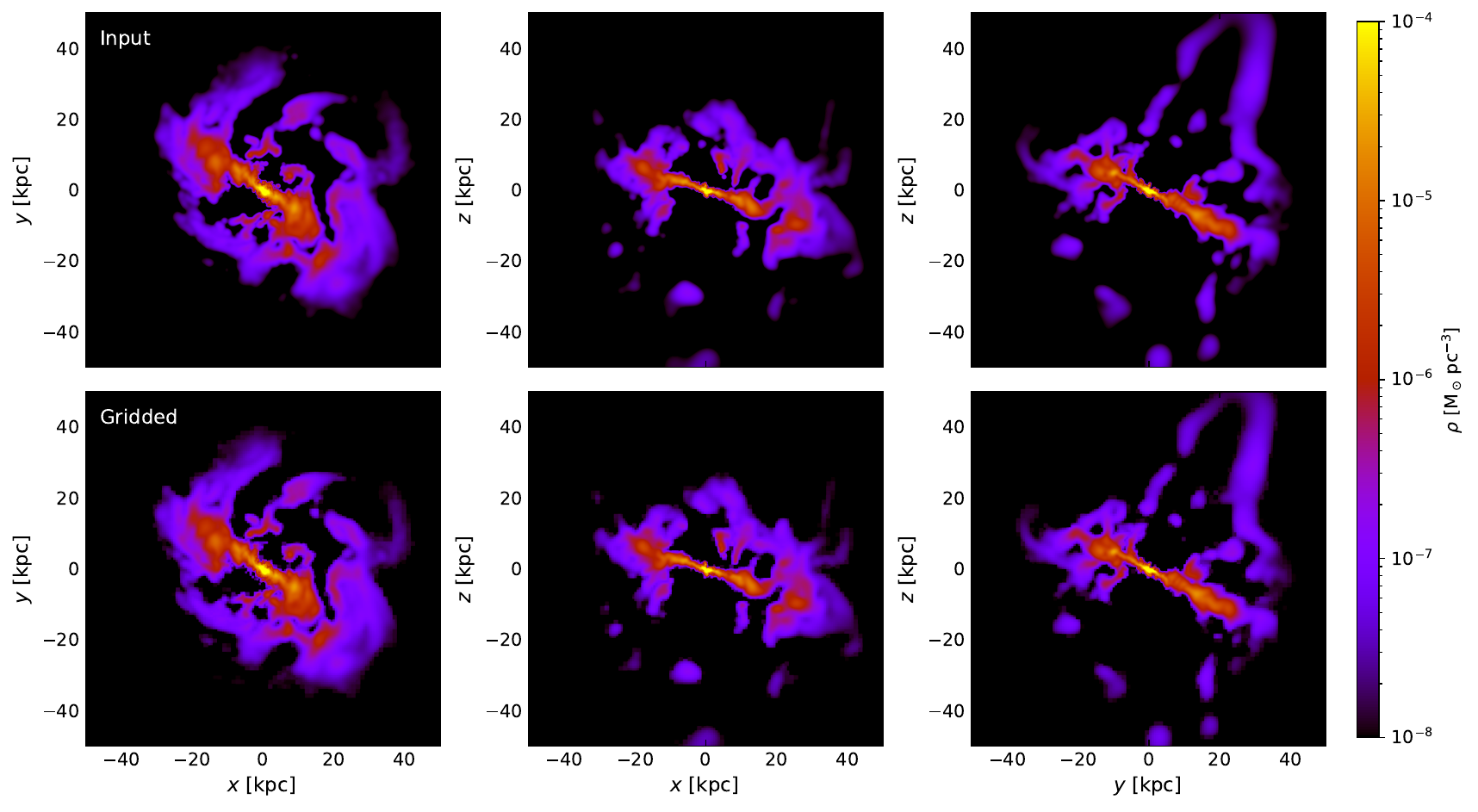}}
     \caption{Input and SKIRT-gridded dust density planar cuts for TNG666210. The upper
     and lower rows show the input and gridded fields, respectively, while the
     columns correspond to the $xy$, $xz$, and $yz$ planes. Only the central
     $100\times100\,{\rm kpc}^2$ region is shown; the full octree domain is
     $[-83,83]^3\,{\rm kpc}^3$.}
      \label{figa1}
\end{figure*}

\begin{figure*}[ht!]
    \centering
    {\includegraphics[width=0.8\textwidth]{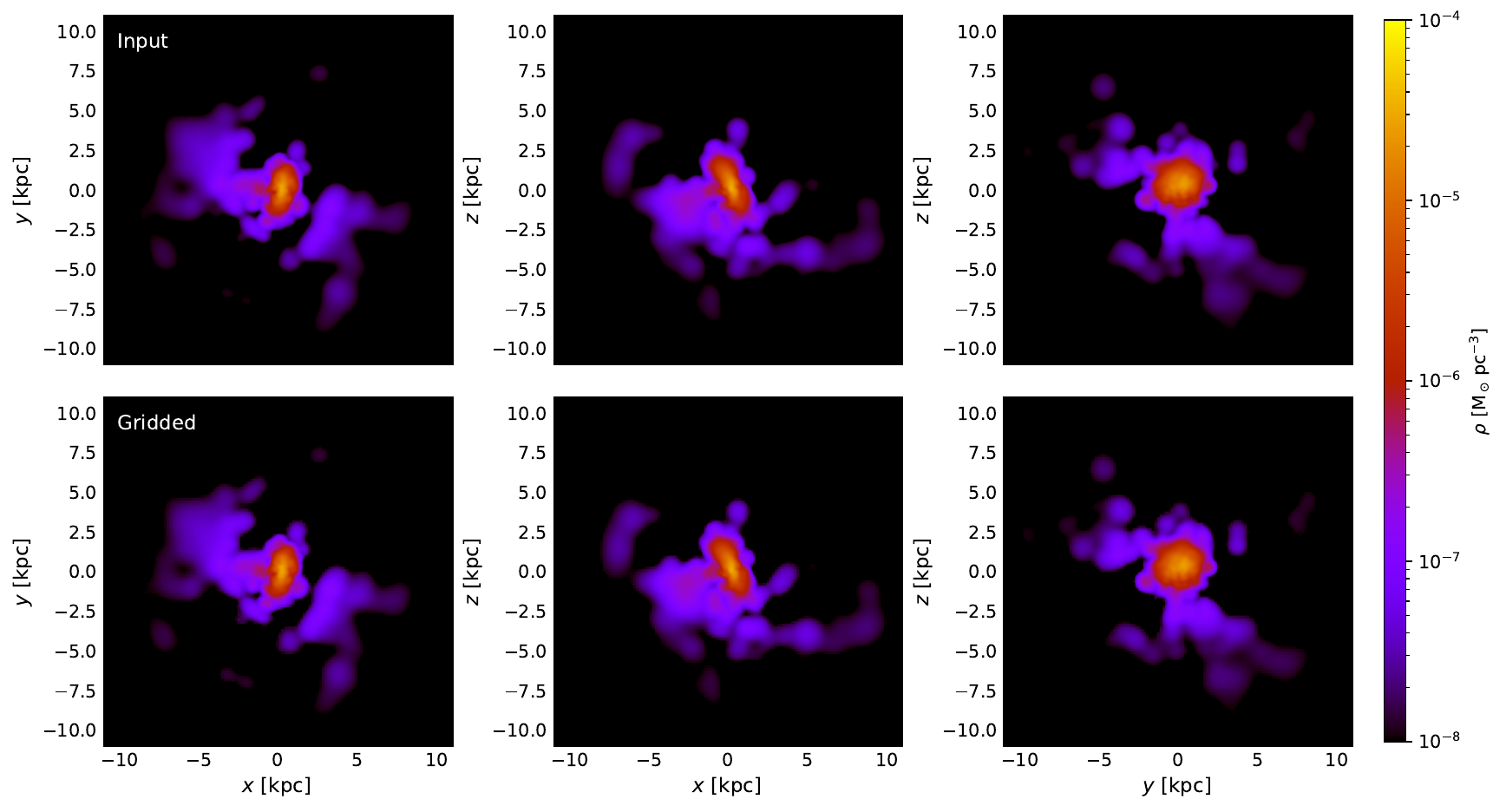}}
     \caption{Same as Fig.~\ref{figa1}, but for TNG860672. Each panel shows the full
$22\times22\,{\rm kpc}^2$ extent of the octree domain.}
      \label{figa2}
\end{figure*}
\end{appendix}
\end{document}